\documentclass[12pt]{article}
\usepackage{amssymb}
\usepackage{amsfonts}
\usepackage{amsmath,amsthm}
\usepackage{subfigure}
\usepackage{graphics,epsfig}
\usepackage{graphicx}
\usepackage{graphicx,epsfig, color }
\oddsidemargin 10mm \numberwithin{equation}{section}
\def\be{\begin{equation}}
\def\ee{\end{equation}}
\begin{document}
\begin{center} {{\bf {Holographic entanglement entropy for small subregions and thermalization of Born-Infeld AdS black holes }}\\
 \vskip 0.50 cm
  {{ H. Ghaffarnejad \footnote{E-mail:
 hghafarnejad@semnan.ac.ir
 } }{ M. Farsam \footnote{E-mail: mhdfarsam@semnan.ac.ir
 } }{ E. Yaraie \footnote{E-mail: eyaraie@semnan.ac.ir
 } }}\vskip 0.2 cm \textit{Faculty of Physics, Semnan
University, 35131-19111, Semnan, Iran }}
\end{center}
\begin{abstract}
Applying the Born-Infeld Anti de Sitter charged black hole metric
 we calculate
holographic entanglement entropy (HEE) by regarding the proposal of
Ryu and Takanayagi.  To do so we assume that time dependence of
the black hole mass and charge to be as step function. Our work is
restricted to small subregions where a
  collapsing null shell dose not penetrate the black holes horizon. To calculate time dependent HEE we
  use perturbation method for small subregions where turning point is much smaller than local equilibrium point of black hole. We choose two shape functions for entangled regions on the boundary which are the strip and the ball regions. There is a saturation time at which the null shell grazes the turning point and the HEE reaches to
  its maximum value.
In general,  this work satisfies
   result of the works presented by Camelio et al and
Zeng et al. We must point out that they used equal time two-point
correlation functions and Wilson loops instead of the entanglement
entropy (EE) as non-local observable to study this
thermalization by applying the numerical method.
\end{abstract}
\section{Introduction}
Duality correspondence between an Anti de Sitter spacetime in bulk
and  boundary conformal field theory (AdS/CFT) is a conjecture
where
 a gravity theory
defined on the AdS bulk space time corresponds to a quantum field
theory defined on its asymptotic boundary region [1,2]. In fact
 the AdS/CFT duality relates quantum physics of strongly coupled
many-body systems to the classical dynamics of a gravitational
model which lives in one higher dimension. It is a suitable way to
understand and to study quantum gravity [3], because one can
argue unknown quantum gravity which is described entirely by a
topological quantum field theory, in which all physical degrees of
freedom are projected onto the boundary.
 The entanglement entropy (EE) or von Neumann entropy of quantum microstates of
boundary CFT behaves as a non-local observable same as the equal
time Wightman two-point correlation functions [4,5] and Wilson
loops [6]. The EE is a very important quantity to understand the
physical nature of thermalization of a non-equilibrium quantum
system [7-17], superconducting phase transition [18-26] and
cosmological singularity [27,28]. Geometrizing EE can be possible by
applying the holographic approach. In this prescription the
EE obtained from the field theory under consideration corresponds
to a minimal surface defined in the bulk geometry which is
anchored to the entangled region.  The holographic approach
presented by Ryu and Takayanagi leads to similarity between the
Bekenstein-Hawking entropy of static AdS black holes and the EE
 of microstates of the dual boundary time-independent
quantum conformal fields [29] (see also [30]). To calculate time
dependent EE the Takayanagi et al proposal is  given in ref. [31]
for time dependent AdS/CFT correspondence where minimal surface
evaluates versus the time until
 equilibrates.
With the above conditions, the minimal surface can penetrate
into the event horizon and gives information from the black hole
structure to the observer far from the horizon without exhibits
the black hole singularity.
 Evolving the EE and the dynamical
process in the bulk is described by a global quantum quench on the
boundary. Actually the
 initial static background in the pure AdS state is perturbed by
  a time-dependent disturbance which is produced by
   injecting
  uniform energy density at
 the initial time. These global quenches on the boundary are modeled by a collapsing null shell of matter which
 initially are located on
 the
boundary and causes to form a black hole at the center of the AdS
 space time [32,33,34]. If  the energy density
 injection  occurs instantaneously,
 then the corresponding disturbance  will be
 sharp
  which  is in accord to  collapse
  of a  very thin shell, while smooth perturbations read
    to specified thickness
     shell.\\
Entanglement evolution obtained from global quantum quench is
studied in [35] for a transverse Ising spin chain in
1+1-dimensional CFT in which the EE grows linearly versus the time
$\Delta S(t)\approx t$ and is proportional to final thermal
entropy density at saturation time $t_{sat}$. The saturation time
is particular time where the system reaches to an equilibrium
state and so EE does not evolve more. Its maximum value is
proportional to the size of entangled region. In general, rate of
EE growth means speed of time evolution of the EE which is defined
by $\Re(t)\sim\frac{d}{dt}\Delta S(t).$ In fact $\Re(t)$ maybe
take on numerical values $\Re(t)\leq1$ which satisfies causality
constraint condition of the holographic system [36,37] while there
maybe are situations for which the causality constraint condition
breaks and maximum value of the HEE velocity raises to some values
bigger that the light velocity. In tsunami picture in which the
evolution of EE is pictured as an inward moving from the boundary
with `tsunami` velocities $ \Re(t)\leq1$ and we can see for the
large subsystems in [14, 51]. The situation is different in small
subsystems which we will consider here. First we define $t_{theq}$
as characteristic time of thermal equilibrium state of the small
subsystem. Now due to
 the size of  the entangled region is much smaller than the
 thermal excitations $t_{sat}\ll t_{theq},$ therefore the interactions counterpart become
 negligible and so the system
takes on its saturate state before the thermal excitations can
affect. \\
Camilo et al studied  holographic thermalization of the
  Born-Infeld AdS black hole in the presence of a chemical potential (the electric charge effect) [38].
   They used equal time two-point correlation functions and expectation values of Wilson loop operators as
   the non-local observable to probe the black hole thermalization. They used numerical method to calculate the dynamical equations and
    the HEE of the system.
   They showed that as the charge and the Born Infeld parameter grow, the thermalization time bears an increase and decrease, respectively.
   They obtained similar result by applying the HEE as the other non-local observable instead of the Wilson loop and two point correlation function.
   They exhibited existence of a phase transition point which is depended to numerical values of the charge $Q$ and the Born-Infeld parameter $b$.
   This point divides the thermalization process into an
   accelerating and a decelerating phase.  As non-local observable
   the two point correlation function and the HEE are used also to study phase transition of the AdS Born-Infeld black hole by Zeng et al [39].
   They obtained that it is happened for
   $bQ>0.5$ but not for $bQ<0.5.$ In the latter case there is a new branch for the infinitesimally
   small black hole so that a pseudo phase transition emerges
   besides the original Hawking-Page phase transition.
   At last they can infer that the phase structure of the
   non-local observable is similar to that of the thermal
   entropy regardless of the size of the boundary region in the
   field theory.
   \\
  Here we explore the bulk geometry by applying the Non-linear Maxwell-Einstein (Born-Infeld) gravity theory
 [40,41]. Then we calculate HEE of a
Born-Infeld charged AdS black hole made from null shell
 collapsing matter. This null shell matter is assumed to be originated from injection of a matter field on the CFT side of the AdS space as suddenly.
 It must be noticed that the time dependence null shell profiles which we consider here, vanishes any
 non-monotonic behaviors during the quench [61], and could get our calculation as simple and lead to a neat result containing
  all of issues which we  seek them.
To do so the
 time dependence for the mass and the
 charge  of the black hole is assumed to have a step function form. This means that the matter source is injected suddenly at the
initial time into the vacuum AdS
 space time to make a black hole.
 To calculate the HEE we should choose a shape function for a subregion on the holographic side. In usual way we choose
the strip and the ball shape for the subregions.
 We calculate the HEE and its evolution rate $\Re(t)$ for different values
 of the dimensionless Born-Infeld coupling constant $b$.\\
Layout of the  paper is as follows. In section 2 we introduce AdS
Einstein-Born-Infeld gravity theory, and corresponding spherically
symmetric static black hole metric solution [40,41,42]. We use
infinite volume limit proposed by Witten [43] to calculate the
effective counterpart of the spherically symmetric static
Born-Infeld black hole metric. It is useful to obtain its
asymptotically AdS form. Then we calculate the corresponding
Hawking temperature. We will use Eddington-Finkelstein coordinates
system to obtain the perturbed metric of the bulk AdS space time
via a collapsing null shell in the presence of nonlinear
electromagnetic field. At last we calculate the location of
turning point $z_t$ with respect to the black hole horizon $z_h$,
mass of the collapsing thin shell $M$, the Born-Infeld coupling
constant $b$ and the AdS radius $L$. For small subregions we have
$z_t\ll z_h$ which  makes us capable to use perturbation method to
calculate the HEE. In section 3 we use Takayanagi et al proposal
to obtain a perturbation series solution for the time dependent
HEE of the Born-Infeld AdS black hole. Section 4 denotes to
conclusion of the work and some outlooks which we consider for the
future work.
\section{AdS-Einstein-Born-Infeld gravity}
Let us we start with a non-linear Maxwell action minimally coupled
with the Einstein-Hilbert gravity action in the presence of a
cosmological constant described in a 4D curved space time as
follows [40,41].
\begin{equation}
S=\int d^4x \sqrt{-g}\bigg[\frac{R_{\mu}^{\mu}-2\Lambda}{16\pi
G}+4b^2\bigg(1-\sqrt{1+\frac{2F}{b^2}}\bigg)\bigg],
\end{equation}
where $R_{\mu}^{\mu}$ is Ricci scalar,
$F=\frac{1}{4}F_{\mu\nu}F^{\mu\nu}$ is the Maxwell electromagnetic
field lagrangian density and $b$ is the Born-Infeld parameter. $G$
is Newton`s coupling constant and $\Lambda$ is the cosmological
constant which relates to radius of the 4D-AdS space time $L$  as
$\Lambda=-\frac{3}{L^2}$ [41]. One can infer for $b\to\infty$ the
last term of the above action leads to
$4b^2[1-\sqrt{1+2F/b^2}]\sim-4F+O(b^{-2})$ and so the total action
 reduces to the well known Einstein-Maxwell
model. Author of the ref. [41] obtained a spherically symmetric
static charged black hole metric of the model (2.1) as follows.
\begin{equation}
ds^2=-f(r)dt^2+\frac{dr^2}{ f(r)}+r^2d\Omega_2^2,
\end{equation}
where $d\Omega_2^2=(d\theta^2+\sin^2\theta d\phi^2)$ denotes to
the unit 2-sphere metric $\mathbb{S}^2$, and $f(r)=1+g(r)$ for
which
\begin{equation}
g(r)=-
\frac{2M}{r}+\frac{r^2}{L^2}+\frac{2b^2r^2}{3}\bigg(1-\sqrt{\frac{Q^2}{b^2r^4}+1}\bigg)
+\frac{4Q^2}{3r^2}~{}_2F_1\bigg(\frac{1}{4},\frac{1}{2},\frac{5}{4};-\frac{Q^2}{b^2r^4}\bigg).
\end{equation}
${}_2F_1(r)$ given in the above metric potential is hypergeometric
function, $M$ and $Q$ are
 constants
 of integral. They have related
to the black hole mass and the electric charge respectively. It is
simple to calculate asymptotically behavior of the metric
potential $f(r)$ at infinity $r\to\infty$ as
$f(r)\sim1-\frac{2M}{r}+\frac{Q^2}{r^2}+\frac{r^2}{L^2}-\frac{Q^4}{20b^2
r^6}$ which leads to a Reissner-Nordstr\"{o}m AdS black hole in
limits $b\to\infty$ [41]. One can obtain nonzero component of the
electromagnetic field of the system as
$F^{rt}=\frac{b}{\sqrt{1+b^2r^4/Q^2}}.$ It is equivalent with the
vector potential gauge field $A_\mu(r)=(A_t(r),0,0,0)$ as
$F^{rt}(r)=-\frac{dA_t(r)}{dr}.$ We can integrate it to obtain
$A_t(r)=\frac{Q}{r}{}_2F_1(\frac{1}{4},\frac{1}{2},\frac{4}{4};-\frac{Q^2}{b^2r^4})-\Phi
$  in which $\Phi$ is a constant of integral. It is really
electrostatic potential difference between the horizon and
infinity of the charged Born-Infeld black hole. $A_t(r)$ is a
gauge field and so we can set $A_t(r_h)=0$ on the black hole
horizon $r_h$ obtained from $f(r_h)=0$ (see [38,41] for more
details.) By applying the latter boundary condition one can infer
\begin{equation}
\Phi(r_h)
=\frac{Q}{r_h}~{}_2F_1\bigg(\frac{1}{4},\frac{1}{2},\frac{5}{4};-\frac{Q^2}{b^2r_h^4}\bigg).
\end{equation}
In the context of the AdS/CFT correspondence there may be exist
some black holes with varied topologies for instance planer
solutions [43] where their topology at the boundary of a 4D-AdS
space-time are $\mathbb{R}^3$ instead of
$\mathbb{R}\times\mathbb{S}^2$. Such a black hole exist only due
to the presence of a negative cosmological constant. To obtain
such a black hole we should usually use finite-volume re-scaling
of the solutions as done in [42] by introducing a dimensionless
parameter $\lambda$ and then obtain its limits at
$\lambda\to\infty$. The latter proposal is well known as "infinite
volume limit" which we will consider it to obtain topologically
deformed form of the metric solution (2.3). It should be reminded
the Born-Infeld parameter $b$ and the AdS radius $L$ is
topological invariant quantities for the metric solution (2.3) but
its other quantities are changed under the transformations
$\lambda$ as follows: $r\rightarrow\lambda^{1/3}r$,
$t\rightarrow-\lambda^{-1/3}t$, $M\rightarrow\lambda M$,
$Q\rightarrow\lambda^{2/3}Q,$
$L^2d\Omega_2^2\rightarrow\lambda^{-2/3}(d\vec{x}\cdot d\vec{x})$
for which $f(r)\to
\lambda^{\frac{2}{3}}\big(\lambda^{-\frac{2}{3}}+g(r)\big).$ After
substituting the latter transformations into the metric solution
(2.2) and taking its limits for $\lambda\to\infty$ we obtain
\begin{equation}
ds^2=-g(r)dt^2+\frac{dr^2}{g(r)}+\frac{r^2}{L^2} d\vec{x}\cdot
d\vec{x},
\end{equation}
where $g(r)$ is given by (2.3) and now the horizon defined by
$g(r_h)=0$ is planner instead of the spherical and so it should be
called as black brane instead of black hole in higher dimensional
gravity models. The Hawking temperature for such a black hole in
the context of AdS/CFT perspective is viewed as the equilibrium
temperature of the dual field theory on the boundary of 4D-AdS
space-time. It is defined by
$T=\frac{1}{4\pi}\frac{dg(r)}{dr}\big|_{r_h}$ which for the metric
equation (2.5) reads [38],
\begin{equation}
T=\frac{r_h}{4\pi}\bigg[2b^2\bigg(1-\sqrt{1+\frac{Q^2}{b^2r_h^4}}\bigg)+\frac{3}{L^2}\bigg].
\end{equation}
In case where the above temperature vanishes $T=0$ then the
charged Born-Infeld black hole
 is called as extremal black hole. It is happened at a particular charge value which its absolute value is given by
\begin{equation}
Q_{max}=\frac{\sqrt{3}r_h^2}{L}\sqrt{1+\frac{3}{4L^2b^2}}.
\end{equation}
In general the above temperature takes on some positive values
$T\geq0$ for which we just have $0<|Q|\leq Q_{max}.$ $Q=0$
corresponds to vanishing vector potential gauge field and
$|Q|=Q_{max}$ corresponds to zero temperature state respectively.
In the latter case the black hole system exhibits with the
thermalization when the HEE difference raises versus the time
because of the thermodynamic equation $\Delta S=\frac{dE}{T}>0.$
In fact for $T>0$ the black hole absorbs the energy from the
environment $\Delta E>0$ and so its HEE difference raises $\Delta
S>0$ and vice versa (see figures 1 and 2).  To study behavior of
the metric solution (2.5) near the AdS boundary, it
 is convenient to introduce a
new `inverse` radial coordinate like $z=L^2/r$ in which the
singularity $r=0$ sits at infinity while AdS boundary stays at
$z=0$.
 Applying
the latter transformation the metric solution (2.5) reads
\begin{equation}
ds^2=\frac{L^2}{z^2}\bigg[-F(z)dt^2+\frac{dz^2}{F(z)}+d\vec{x}\cdot
d\vec{x}\bigg],
\end{equation}
where,
\begin{gather}
 \nonumber F(z)=\frac{z^2}{L^2}g(L^2/z)=1-\frac{2Mz^3}{{
 L^4}}
 +\frac{4Q^2z^4}{3L^{6}}~{}_2F_1\bigg(\frac{1}{4},\frac{1}{2},\frac{5}{4};
 -\frac{Q^2z^4}{b^2L^8}\bigg)\\
+\frac{2b^2L^{2}}{3}\bigg(1-\sqrt{\frac{Q^2z^4}{b^2L^8}+1}\bigg)
\end{gather}
for which $F(z)\to1$ near the AdS boundary $z=0.$ This means that
near the AdS boundary the metric (2.8) leads to a conformaly flat
form with conformal factor $\frac{L^2}{z^2}.$ In order to avoid
the coordinate singularity at $r=r_h$ we should use the
Eddington-Finkelstein coordinate system $(v,z,x_i; i=3,4)$ with
$d\upsilon=dt-dz/F(z)$ to re-write the metric solution (2.8) as
follows.
\begin{equation}
ds^2=\frac{L^2}{z^2}\bigg[-F(z)d\upsilon^2-2d\upsilon
dz+d\vec{x}\cdot d\vec{x}\bigg].
\end{equation}
 Up to the term of conformal
factor $L^2/z^2,$ the above metric is usually called as Vaidya
space time for time dependent metric $F(z,v)$. It describes space
time metric from point of view of an accelerating observer moving
toward the black hole center. In general the Hawking temperature
of black holes destruct its background metric which is well known
as backreaction problem (see introduction section in ref. [45]).
In the particle physics perspective one can infer some interacting
quantum matter fields fluctuations create particles-antiparticles
 near the horizon. Stress tensor of these created particles causes to shrink the horizon and so the horizon maybe eliminated finally and so one
can exhibit with an essential question as follows. What is final
state of an evaporating black hole? In usual way it is still an
open problem and the backreaction equation
$G_{\mu\nu}=<\widehat{T}^{quantum~field}_{\mu\nu}>_{renormalized}$
is solved analytically just for 2D black holes (see [45,46] and
references therein). As a future work we encourage to study HEE of
this black hole by regarding the mentioned above backreaction
problem. With this correspondence the Bekenstein-Hawking entropy
of an evaporating AdS black hole become equivalent with the EE of
quantum microstates from CFT side. It is done when the minimal
surface wrap the horizon [29]. In fact holography provides a
geometric intuition for which local operators such as expectation
values of the energy momentum tensor of the quantum matter fields
$<\widehat{T}^{quantum~field}_{\mu\nu}>_{renormalized}$ are not
sensitive to the thermalization process. They are only sensitive
to phenomena happening near the AdS boundary. To probe the global
process of thermalization one needs to consider extended non-local
observables such as two point correlation functions and
expectation values of rectangular Wilson loops and  EE of boundary
regions which have well known holographic descriptions in the dual
AdS bulk perspective (see [38,42] for more discussions ). In
general one
 use usually a collapsing  thin shell to study dynamics of an
evaporating quantum black hole. If the interacting quantum matter
fields become charge-less then the electric charge of the quantum
black hole will be invariant while its mass evaporates and so we
have to put $M\to M(v)$ in the metric equation (2.11). If the
quantum matter fields have electric charge (for instance complex
quantum scalar fields) the quantum black hole electric charge
varies for which we should  replace $Q\to Q(v)$ together with the
$M\to M(v)$. Authors of the paper [42] are used samples
$M(\upsilon)=M(1+\tanh\frac{\upsilon}{\upsilon_0})/2$ and
$Q(\upsilon)=Q(1+\tanh\frac{\upsilon}{\upsilon_0})/2$, to study
holographic thermalization of a charged-Born-Infeld quantum black
brane (2.11). It is clear that for  $\upsilon\to-\infty$ we have
pure AdS with $M(-\infty)=0=Q(-\infty)$ while for
$\upsilon\to+\infty$ we have $M(+\infty)=M,Q(+\infty)=Q$.
$\upsilon_0$ is a constant and shows the finite thickness of the
shell. The Smooth form for the above mentioned sample
$M(\upsilon), Q(\upsilon)$ shows that they are applicable for the
numerical
  analysis. For
 a shock wave which is made from zero thickness shell of charged
 matter suddenly forming at $\upsilon=0$ we can usually use a step function $Q(\upsilon)=Q\theta(\upsilon)$ and
 $M(\upsilon)=M\theta(\upsilon)$ (see section 3.3.1 in ref. [47])
  which is applicable for analytical calculations. They
 are asymptotic behavior of the functions $M(\upsilon)=M(1+\tanh\frac{\upsilon}{\upsilon_0})/2$ and
$Q(\upsilon)=Q(1+\tanh\frac{\upsilon}{\upsilon_0})/2$, at
$\upsilon_0\to0$ which we will consider in this paper (see [38]
for more discussions). To obtain time dependent form of the metric
solution (2.11) it is convenient to evaluate mass and charge of
the black hole versus the corresponding Hawking temperature (2.6)
as follows.\\ Applying the inverse transformation
$z=\frac{L^2}{r}$ for the event horizon $r_h$ obtained from
$g(r_h)=0$ one can determine location of the event horizon in the
Eddington-Finkelstein coordinate system as $z_h=\frac{L^2}{r_h}.$
In fact $z_h$ is determined versus the AdS Born Infeld black hole
parameters $M,Q,b,L$ by solving $F(z)=0$ given by (2.9) as
follows.
\begin{equation}1-\frac{2Mz_h^3}{{
 L^4}}
 +\frac{4Q^2z_h^4}{3L^{6}}~{}_2F_1\bigg(\frac{1}{4},\frac{1}{2},\frac{5}{4};
 -\frac{Q^2z_h^4}{b^2L^8}\bigg)
+\frac{2b^2L^{2}}{3}\bigg(1-\sqrt{\frac{Q^2z_h^4}{b^2L^8}+1}\bigg)=0.\end{equation}
Substituting $r_h=\frac{L^2}{z_h}$ into the relation (2.6) we can
infer:
\begin{equation}T(M,Q,b,L)=\frac{3+2b^2L^2(1-\sqrt{1+Q^2z_h^4/b^2L^8})}{4\pi z_h}.
\end{equation} In the collapsing thin shell model, we assume  $\mathcal{R}(t)>l_{eq}$ to be the size of an evolving spacelike surface $(\Sigma,t)$
at a time `$t$, where $l_{eq}$ is its size at in equilibrium
state. The corresponding bulk extremal surface $\gamma_\Sigma(t)$
can be parameterized as $(z_t(t),\upsilon_t(t))$ versus the time
parameter `t`. There are usually  exist multiple extremal surfaces
for a given $(\Sigma,t)$ for which we should choose the one with
smallest area. In general there is not a simple relation between
$(\mathcal{R},t)$ and $(z_t,\upsilon_t)$ for which we must be
solve the full equations of motion for entangled area
$\gamma_{\Sigma}(t).$ In a penrose diagram of the collapsing
system $\gamma_{\Sigma}(t)$ trace out a curve
$(z_t(t),\upsilon_t(t))$ for a given $(\Sigma,t)$ by varying the
time parameter `t`(See figs. 2 and 3 in ref. [14]). These figures
show that the location of event horizon $(z_h,\upsilon_h)$ is a
local equilibrium scale and so for
$z_t=\frac{L^2}{R(r)},z_h=\frac{L^2}{r_h}$ one can infer
$\frac{z_t}{z_h}<1.$ Now we are in a position to obtain time
dependent form of the metric (2.11) for a collapsing thin shell
with mass $M(\upsilon)$ and the charge $Q(\upsilon)$ as follows.
We will try to obtain this time behavior by a perturbation method
presented in [47] in which the on shell action is expanded with
respect to a dimensionless perturbation parameter. In our case
this parameter would be $\epsilon\equiv\frac{z_t}{z_h}\ll1$ as it
mentioned in above. By expanding the Lagrangian density and
embedding functions around the vacuum case one can obtain the
on-shell action as follows.
\begin{eqnarray}
\nonumber S_{on-shell}[\varphi(z)]=\int_{on-shell} dz\mathcal{L}[\varphi(z),\epsilon]\\=\int
dz\mathcal{L}^{(0)}[\varphi^{(0)}(z)]+\epsilon\int dz\mathcal{L}^{(1)}[\varphi^{(0)}(z)]+\mathcal{O}(\epsilon^2),
\end{eqnarray}
in which $\mathcal{L}$ is the Lagrangian density and $\varphi(z)$
stands for embedding functions which could be obtained by solving
the Euler Lagrange equations. Neglecting higher order terms of the
action functional (2.13) we can infer that leading order term of
the embedding functions, $\varphi^{(0)}(z)$ is just enough to
obtain on-shell action. It is enough to compatible with
observation. To have a perturbation solution for the model it is
better to re-write the functional with respect to the orders
of $z_t/z_h$ and expand $F(z)$ defined by (2.9) as well as embedding functions order by order.\\
But before doing that it is convenient to evaluate mass and charge
of black hole with respect to the Hawking temperature. Since
Hawking temperature contains event horizon, so $z_h$ appears into
the terms and perturbation parameter can be extracted easily.
Putting $L=1$ for convenient and using (2.12) for the temperature
and $F(z_h)=0$ for the event horizon we get
\begin{equation}
M=\frac{\mu}{z_h^3},~~~\text{and}~~~Q^2=\frac{\eta}{4b^2z_h^4},
\end{equation}
with
\begin{gather}
\nonumber\eta=4\pi Tz_h(4\pi Tz_h-4b^2-6)+12b^2+9,\\
\mu=\frac{1}{2}+\frac{\eta}{6b^2}~{}_2F_1\bigg(\frac{1}{4},\frac{1}{2},\frac{5}{4};-\frac{\eta}{4b^4}\bigg)
+\frac{b^2}{3}\bigg(1-\frac{1}{2}\sqrt{4+\frac{\eta}{b^4}}~\bigg).
\end{gather}
 Thus by these new parameters which are functions with respect to
the event horizon and the Hawking temperature, $F(\upsilon,z)$ can
be rewritten as
\begin{gather}
\nonumber F(\upsilon,z)=1-2\mu(\upsilon)\bigg(\frac{z}{z_h}\bigg)^3+\frac{\eta(\upsilon)}{3b^2}~{}_2F_1\bigg(\frac{1}{4},
\frac{1}{2},\frac{5}{4};-\frac{\eta(\upsilon)}{4b^4}\bigg(\frac{z}{z_h}\bigg)^4\bigg)\bigg(\frac{z}{z_h}\bigg)^4\\
+\frac{2b^2}{3}\bigg(1-\sqrt{1+\frac{\eta(\upsilon)}{4b^4}\bigg(\frac{z}{z_h}\bigg)^4}~\bigg).
\end{gather}
Time dependence of $\eta(\upsilon)$ and $\mu(\upsilon)$ comes from time evolution of the mass and charge of black hole. We
can choose this dependence in the form of step functions for simplicity as $\mu(\upsilon)=\mu~\theta(\upsilon)$ and
 $\eta(\upsilon)=\eta~\theta(\upsilon)$, and at last we obtain $F(z)$ again but with respect to $\eta$ and $\mu$ instead of the mass and charge.\\
Now by attention to what is mentioned before if $z_t\ll z_h$ and since always $z\leqslant z_t$ therefore $(z/z_h)\ll1$, and by keeping two
first term from expansion of (2.16) we get to
\begin{equation}
F(\upsilon,z)=1-2\mu~\theta(\upsilon)\bigg(\frac{z}{z_h}\bigg)^3+\frac{\eta~\theta(\upsilon)}{4b^2}\bigg(\frac{z}{z_h}\bigg)^4+\mathcal{O}
\bigg(\frac{z}{z_h}\bigg)^8.
\end{equation}
Zero order term in the above equation as expected represents pure AdS solution which is static and will not
 change over time. We can take $\epsilon\equiv(z_t/z_h)^3$ as dimensionless perturbation parameter. The next term is in
 order of
 $(z_t/z_h)^4=\epsilon^{\frac{4}{3}}$ which is located before the term $\epsilon^2$. In fact it is still first order and without loss of generality it can be rewritten as $(z^4/z_hz_t^3)\epsilon$. Therefore we must treat with this term as the first order and use exactly the same embedding functions of pure AdS in the first order approximation. \\

\section{Time dependent HEE}
According to the  Ryu and Takayanagi proposal [29] in the context
of AdS/CFT correspondence, the EE of a region on the boundary is
defined by
\begin{equation}
S_{EE}=\frac{1}{4G_N^{(d)}}\int_{\Sigma}d\zeta^{d-2}\sqrt{h}=
\frac{1}{4G_N^{(4)}}\int_{\Sigma}dz\bigg(\mathcal{L}^{(0)}+\epsilon\theta(\upsilon)\mathcal{L}^{(1)}\bigg),
\end{equation}
where $\Sigma$ is the minimal area surface with $d$-dimension in
the bulk which is bounded to the CFT boundary  and $h$ is absolute
value of determinant of induced metric defined on $\Sigma.$ The
spatial coordinates $\zeta^i$ with $i=1,2,\cdots,d-2,$ are
world-volume coordinates defined on the $(d-2)-$ dimensional
surface (see for instance [49] and references
therein) which are integrated in the last term and put into the functionals. \\
For dynamical gravitational systems where a time dependent AdS/CFT
correspondence exists we should calculate (3.1) on extremal
surface $\Sigma(t)$ which in general is different with the minimal
surface defined on the static space times. An arbitrary time
dependent spatial surface becomes extremized by vanishing scalar of
extrinsic curvature $K=\nabla_an^a$ where $n^a$ is unit light-like
vector field defined at each point on $\Sigma(t)$ [31].
  Fortunately authors of the ref.
[31] proved that for a particular Vaidya form background metric,
the extrinsic curvature scalar $K$ vanishes trivially by
substituting the equation of motion of the embedding functions
into it (see section 6.3 in ref. [31]). Euler-Lagrange equations
of the embedding functions are obtained by varying (3.1) with
respect to the induced metric fields. The latter proposal helps us
to understand the time evolution of EE which corresponds to
collapsing a null shell from the boundary to form a black hole. \\
Let us start with time dependent form of the metric solution
(2.10) in which $F(z)$ is replaced with $F(\upsilon,z)$ such that
\begin{equation}
ds^2=\frac{L^2}{z^2}\bigg[-F(\upsilon,z)d\upsilon^2-2d\upsilon
dz+d\vec{x}\cdot d\vec{x}\bigg].
\end{equation} where  $d\vec{x}\cdot d\vec{x}=g_{ij}dx^idx^j$ (with ${i,j}=3,4$) could be
defined for any entangled region on the boundary. Here we will choose holographic
region $A(t)$ to be half-cylinder-like and himsphere-like which
will be viewed as strip and circle respectively on the holographic
side (see figure 2 in ref. [29] and figure 4 in ref. [47]).

\subsection{The strip region}
For the strip region we set $\{x,y\}$ to be
 spatial coordinates on the space-like boundary
 hyper-surface $\Sigma(t)$ with corresponding 2-metric $d\vec{x}\cdot d\vec{x}=dx^2+dy^2.$ We assume the
 strip
 is extended along $y$ direction such that: $\{x\in (-\frac{\ell}{2},\frac{\ell}{2}), y\in (0,D>>\ell)\}.$
 The half-cylinder-like extremal surface area and its strip
  is invariant under the translation in $y$-direction.
  In the latter case one can infer shape function of the
 half-cylinder-like will be $(x(z),\upsilon(z))$ in which we choose $z$ to be the holographic coordinate and $\{z,y\}$ are placed  on the holographic
 side. Other boundary conditions of these embedding functions are $\upsilon(0)=t$ and $x(z_t)=0$. The induced metric on the half
 -cylinder-like reads
 \begin{equation}
ds^2=h_{\mu\nu}dx^{\mu}dx^{\nu}=h_{zz}dz^2+h_{yy}dy^2
\end{equation}
with
\begin{equation} h_{yy}(z)=\frac{L^2}{z^2},
~~~h_{zz}(\upsilon,z)=\frac{L^2}{z^2}\bigg(x^{'2}-2\upsilon^{'}-F(\upsilon,z)\upsilon^{'2}\bigg)\end{equation}
in which $x^{'}=\frac{dx}{dz}$ and
$\upsilon^{\prime}=\frac{d\upsilon}{dz}$ should be calculated from
the embedding functions $x(z),\upsilon(z).$ They can be determined
by solving the corresponding Euler-Lagrange equation. The
Euler-Lagrange equations are obtained by varying the EE with
respect to the fields $x(z)$ and $v(z)$ and setting their values
with zero. To do so we calculate the EE defined by (3.1) for which
determinant of the induced metric (3.4) is
\begin{equation} h=\det h_{\mu\nu}=h_{yy}h_{zz}.\end{equation}
Setting $L=1$ and applying (3.4), (3.5) and (2.17) for the action
functional (3.1) we obtain leading order term of Lagrangian
density for the strip region as follows.
\begin{equation}
\mathcal{L}^{(0)}=\frac{A_{\Sigma}}{z^2}\sqrt{x^{\prime
2}-2\upsilon^{\prime}-\upsilon^{\prime 2}}.
\end{equation}
It is un-perturbed counterpart which obtained at $t-constant$
slices spacetime with $F(\upsilon,z)=1$, and
\begin{equation}
\mathcal{L}^{(1)}=\frac{A_{\Sigma}\upsilon^{\prime 2}(z/z_t)^3}{z{
\sqrt{x^{\prime 2}-2\upsilon^{\prime}-\upsilon^{\prime 2}}}}\bigg(\frac{\mu}{z}-\frac{\eta}{8b^2z_h}\bigg),
\end{equation}
corresponds to the first order term. In the above equations
$A_{\Sigma}=\int_{\Sigma}dy=2D$. Furthermore, we can expand the
embedding functions $\upsilon(z)$ and $x(z)$ for small $\epsilon$.
Achieving this goal we start with $d\upsilon=dt-dz/F(z)$, and by
using (2.17) we obtain

\begin{equation}
\upsilon=t-z-\bigg(\frac{\mu}{2z_t^3}~ z^4\bigg)\epsilon+\mathcal{O}(\epsilon^2)~~~~~~\text{for   }\upsilon>0.
\end{equation}
From on-shell expansion (2.13) we remember that the zero order
term of embedding functions is enough for our purpose, so we can
consider
\begin{equation}
\upsilon^{(0)}(z)=t-z.
\end{equation}
By this approximation $\upsilon^{(0)\prime}=-1$, in which as mentioned before the prime is derivative with respect to $z$. Plugging
 this value leads us to the un-perturbed counterpart $\mathcal{L}^{(0)}$ which does not contain $x(z)$ explicitly,
 and so has a constant of motion with respect to the holographic direction $z$.
  By applying boundary condition $x^{(0)\prime}_{(z=0)}\rightarrow\infty$ ( or equivalently $z^{(0)\prime}_{(x=0)}=0$) we yield,
\begin{equation}
x^{(0)\prime}(z)=\pm\frac{\big(\frac{z}{z_t}\big)^2}{\sqrt{1-\big(\frac{z}{z_t}\big)^4}}.
\end{equation}
Since the evolution of $x(z)$ and $z$ are inversely related
($\frac{dx}{dz}<0$), so without loss of generality we can consider
minus sign and get the zero order term of embedding function as
follows.
\begin{equation}
x^{(0)}(z)=\frac{\ell}{2}-\frac{z_t}{3}\bigg(\frac{z}{z_t}\bigg)^3{}_2F_1\bigg(\frac{1}{2},\frac{3}{4},\frac{7}{4};\bigg(\frac{z}{z_t}\bigg)^4\bigg).
\end{equation}
The above equation  which is come from the pure AdS solution [50]
is obtained by integrating (3.10). Setting the boundary condition
$x(z_t)=0$ for (3.11) we could get a relationship between the size
of the strip and turning point as
\begin{equation}
\ell=\frac{2\sqrt{\pi}~\Gamma(\frac{3}{4})}{\Gamma(\frac{1}{4})}~z_t.
\end{equation}
The expansion of (3.1) separates the time evolution of EE in two
parts in which first part is a static solution in pure AdS
spacetime, while the second one is time-dependant solution
affected by black hole formation process. Namely
\begin{equation}
 S(t)=S_{vac}+S^{(1)}(t)+\cdots
\end{equation}
in which dots implies higher order perturbation terms which can be
ignored  here because of their weak affects. The first term is the
vacuum entropy that is constant during the process of black hole
formation such that
\begin{equation}
S_{vac}=\frac{1}{4G_N^{(4)}}\int_{\delta}^{z_t} dz\mathcal{L}^{(0)}[x^{(0)}(z),\upsilon^{(0)}(z)]=\frac{A_{\Sigma}}{4
G_N^{(4)}}\bigg[\frac{1}{\delta}-\frac{2\pi}{\ell}\bigg(\frac{\Gamma(\frac{3}{4})}{\Gamma(\frac{1}{4})}\bigg)^2\bigg],
\end{equation}
where $\delta$ denotes to high energy cut-off scale for the
holographic direction. The second term which appears after
starting the black hole formation in $\upsilon>0$ would be
\begin{gather}
\nonumber S^{(1)}(t)=\frac{1}{4G_N^{(4)}}\int_{0}^{z_t} dz~\epsilon\theta(\upsilon)\mathcal{L}^{(1)}[x^{(0)}(z),\upsilon^{(0)}(z)]\\
=\frac{\mu A_{\Sigma}}{4G_N^{(4)}z_h^3}\int_{0}^{z_t}dz \theta(\upsilon)z\sqrt{1-\big(\frac{z}{z_t}\big)^4}
-\frac{\eta A_{\Sigma}}{32b^2G_N^{(4)}z_h^4}\int_{0}^{z_t}dz \theta(\upsilon) z^2\sqrt{1-\big(\frac{z}{z_t}\big)^4}.
\end{gather}
Since in this work we are only interested in time-dependant behavior of HEE, so we just consider $\Delta S(t)=S(t)-S_{vac}=S^{(1)}(t)$.\\
To solve (3.15) we replace $dz$ with $d\upsilon$ which for
$\upsilon(>0)$ we have $\theta(\upsilon)=1$ and it turns to the
following form.
\begin{eqnarray}
\nonumber S^{(1)}(t)=\frac{\mu A_{\Sigma}}{4G_N^{(4)}z_h^3}\int_{t-z_t}^{t}d\upsilon (t-\upsilon)\sqrt{1-\big(\frac{t-\upsilon}{z_t}\big)^4}\\
-\frac{\eta A_{\Sigma}}{32b^2G_N^{(4)}z_h^4}\int_{t-z_t}^{t}d\upsilon (t-\upsilon)^2\sqrt{1-\big(\frac{t-\upsilon}{z_t}\big)^4}.
\end{eqnarray}
Now according to the limit of above integral, two situations may be arisen. Time dependant behavior of the EE
 is different before and after the specific time scale appropriate to $z_t$ which is called the saturation time and is happened
  when the null shell grazes turning point; in the other words, at saturation time $\upsilon(z_t)=0$ which from (3.9) leads
  to $t_{sat}=z_t$. From (3.12) turning point is appropriate with the size of entangled region and it could be
   concluded that for larger size of the region, the saturation time will be increased. Now, with this saturation time the following two cases
   can be studied:\\
$(a)$ If $t<t_{sat}$ then the range of the integral varies from
zero to $t$ and the final result is a time dependent function such
that
\begin{eqnarray}
\nonumber\Delta S(t<t_{sat})=\frac{\mu A_{\Sigma}}{16G_N^{(4)}z_h^3}t^2\bigg[\sqrt{1-\bigg(
\frac{t}{z_t}\bigg)^4}+{}_2F_{1}\bigg(\frac{1}{2},\frac{1}{2},\frac{3}{2};\bigg(\frac{t}{z_t}\bigg)^4\bigg)\bigg]\\
-\frac{\eta A_{\Sigma}}{160b^2G_N^{(4)}z_h^4}t^3\bigg[\sqrt{1-\bigg(\frac{t}{z_t}\bigg)^4}+\frac{2}{3}~{}_2F_{1}
\bigg(\frac{1}{2},\frac{3}{4},\frac{7}{4};\bigg(\frac{t}{z_t}\bigg)^4\bigg)\bigg].
\end{eqnarray}
$(b)$ If $t>t_{sat}$ then the evolution process which started at
$t=0$ only lasts until the saturation time, and after that doesn't
change anymore. We can simply put $t=t_{sat}$ in (3.17) and get to
\begin{equation}
\Delta S(t>t_{sat})=\frac{\mu\pi A_{\Sigma}z_t^2}{32G_N^{(4)}z_h^3}+\frac{\eta A_{\Sigma}\sqrt{\pi}
z_t^3}{320b^2G_N^{(4)}z_h^4}\frac{\Gamma(-\frac{1}{4})}{\Gamma(\frac{1}{4})}.
\end{equation}
For a schematic study we plotted the evolution of EE for the strip region in figure $(1.a)$ and $(1.b)$ for some values of electric charge and
Born-Infeld parameter. As we can see, this evolution before saturation time is curve-like and EE starts growing as soon as the null
 shell begins to collapse. As it can be observed and we will show later,
 for the initial times this evolution is quadratic and has a parabola shape with increasing gradient,
 but by passing time and for the middle times this evolution would be more linear with almost constant gradient.
 Some times just before saturation time this linear-like behavior is returned again to parabola-like shape but this time with decreasing gradient.
 So the evolution of EE behaves as $O(t^3)$, and with a "reflection point" in the middle of the linear-like phase before reaching to the saturation
  time. At the saturation time, this growth ceases and after that the value of time-dependant part of EE will be fix at a constant value which we call
   "saturation entropy" defined in (3.18) as $\Delta S_{sat}=\Delta S(t>t_{sat})$.
   We can see that for a fixed charge, the evolution of EE varies faster due to sharper gradients in figure $(1.a,b)$
    by increasing
 the Born-Infeld parameter. So the saturation value of EE raises by increasing
   $b$.\\
In figure 2 we plotted the evolution of EE for fixed $b$ and different electric charges. As it mentioned before, when
$b\rightarrow\infty$ we will have a typical AdS solution. In figure $(2.b)$ we see that when $b\to \infty$
then the saturation entropy leads to a definite value labeled by $\Delta S_{AdS}$ which corresponds to AdS-RN solution.
In the same figure the evolution plotted for un-charged case for any value of $b$ which showed by red line and as we
expected has the lowest value compared to non-zero charge cases.  \\
There is a dimensionless quantity which is useful to study the EE
evolution. This quantity represents instantaneous rate of the
growth by factorizing the aspects of the system such as the size
of the region or total number of degrees of freedom. A system with
the bigger size has more degrees of the freedom which leads to
faster speed of the growth for EE. This rate of entanglement
growth is defined by [14,51],
\begin{equation}
\Re(t)=\frac{1}{s_{sat}A_{\Sigma}}\frac{d(\Delta S(t))}{dt},
\end{equation}
where $s_{sat}=\Delta S_{sat}(t)/V_A$ is the equilibrium entropy density of the system which happened
after saturation time and $V_A$ is the volume of the entangled region $A$. Since as it mentioned before $A_{\Sigma}=2D$ and the volume of region
 must be $V_A=D\ell$, we have
\begin{equation}
\Re(t)=\frac{\ell}{2\Delta S_{sat}}\frac{d(\Delta S(t))}{dt}.
\end{equation}
By noticing that $\Delta S_{sat}$ is time-independent, we reach to a statement which is a time dependent
function and independent of all other quantities except $z_t$ (or $\ell$). On the other words $\Re(t)$ is independent of the state whereas
in large subsystems which is studied in refs. [14,51], we can see state dependence situation.
In figure (3.a) the rate of this function is plotted for $Q=1$ in which by varying the Born-Infeld parameter
 we observe negligible changes and approximately all diagrams are similar same.
 But the interesting situation happens when we plot this function for $Q=5$ for which by varying $b$,
 changing in diagrams would be significant.
 As we can see in (3.b) for the strip region that $\Re_{max}$ decreases by decreasing Born-Infeld parameter, but for $b<0.01$
 it grows up until for $b=0.0025$ which reduces to the speed of light. It is plotted by dot-gray line.
 As expected by more decreasing of the Born-Infeld parameter, $\Re_{max}$ exceeds the speed of light which
 does not have any similar situation in the large subsystems at all.\\
\subsection{The ball region}
For a $2$-dimensional ball region on CFT side with radius $a$ we
can define a radial coordinate on the boundary as $r_b\leqslant
a$, and $d\vec{x}^{~2}=dr_b^2+r_b^2d\varphi_b^2$ in which
$0<\varphi_b<2\pi$. It must be noticed that we use $r_b$ and
$\varphi_b$ for polar coordinates on the boundary to avoid making
mistake with bulk coordinates. On the other side extremal surface
is invariant under rotation and could be parameterized by
embedding functions $r(z)$ and $\upsilon(z)$ which satisfy the
following boundary conditions.
\begin{equation}
r_b(0)=a,~~~\upsilon(0)=t,~~~r_b(z_t)=0,
\end{equation}
where $z_t$ denotes to the turning point in case of the  ball region.\\
The induced metric for this ball region reads
\begin{equation}ds^2=h_{zz}dz^2+h_{\varphi\varphi}d\varphi^2
\end{equation} in which
\begin{equation}h_{\varphi\varphi}=\frac{L^2r_b^2}{z^2},~~~
 h_{zz}=\frac{L^2}{z^2}[r_b^{\prime2}-2\upsilon^{\prime}-F(\upsilon,z)\upsilon^{\prime2}]\end{equation}
 with determinant
\begin{equation}
h=\det h_{\mu\nu}=h_{zz}h_{\varphi\varphi}
\end{equation}
where prime denotes to differentiation with respect to $z$
coordinate. By attention to the above considerations and by
setting $L=1$ we obtain series components of lagrangian functional
(3.1) for the ball region as follows.
\begin{equation}
\mathcal{L}^{(0)}=\frac{A_{\Sigma}r_b}{a z^2}\sqrt{r_b^{\prime 2}-2\upsilon^{\prime}-\upsilon^{\prime 2}},
\end{equation}
and,
\begin{equation}
\mathcal{L}^{(1)}=\frac{A_{\Sigma}r_b}{a z}\frac{\upsilon^{\prime 2}(z/z_t)^3}{\sqrt{r_b^{\prime 2}-2\upsilon^{\prime}-\upsilon^{\prime 2}}}
\bigg(\frac{\mu}{z}-\frac{\eta}{8b^2z_h}\bigg),
\end{equation}
where $A_{\Sigma}=\int_{\Sigma}r_bd\varphi_b$, for which $A_{\Sigma}$ is the area of 2-dimension ball (or disk) with radius $r_b$. So by these consideration we get to $A_{\Sigma}=2\pi a$.\\
Like the strip case, we solve Euler-Lagrange equation obtained
from un-perturbed lagrangian functional (3.25) to obtain equation
of motions of embedding functions $r_b^{(0)}(z)$ and
$\upsilon^{(0)}(z)$. But since in contrary with the strip case,
Lagrangian functional includes the embedding function $r_b(z)$ and
its derivative, so there is not any constant of motion and we have
to solve the equations explicitly. Because these equations are bit
complicated, so the analytical solution gets us into trouble. For
these complicated equations an acceptable suggestion for
$r^{(0)}_b(z)$ could be [50],
\begin{equation}
r_b^{(0)}(z)=\sqrt{z_t^2-z^2},
\end{equation}
for which the boundary condition $r^{(0)}_b(z_t)=0$ (lies at the center of the disk) satisfies well.
By another boundary condition we can obtain the relation between turning point and the region size,
\begin{equation}
r_b^{(0)}(z=0)=a~~\Rightarrow~~z_t=a.
\end{equation}
This relation says that the deepest point of the extremal surface has the same length with the disk radius
 and therefore indicates our extremal surface is ball-shaped.
 In addition, derivative of (3.27) with respect to $z$ for $z\ll z_t$ leads to the
 following equation.
\begin{equation}
r_b^{(0)\prime}=-\frac{\big(\frac{z}{z_t}\big)}{\sqrt{1-\big(\frac{z}{z_t}\big)^2}}.
\end{equation}
For another embedding function, $\upsilon^{(0)}(z)$, situation doesn't changed and therefore (3.9) is valid and so $\upsilon^{(0)\prime}=-1$.\\
Similar to the strip case we are interested to time-dependant part
of EE which lies inside of the collapsing null shell with
$\upsilon>0$, so we only compute the following part which is the
difference between total EE and the pure AdS part of extremal
surface obtained from (3.1) as follows.
\begin{gather}
\nonumber S^{(1)}(t)=\frac{1}{4G_N^{(4)}}\int_{0}^{z_t} dz~\epsilon\theta(\upsilon)\mathcal{L}^{(1)}[r_b^{(0)}(z),\upsilon^{(0)}(z)]\\
=\frac{\mu A_{\Sigma}}{4G_N^{(4)}z_h^3}\int_{0}^{z_t}dz \theta(\upsilon)z\bigg(1-\big(\frac{z}{z_t}\big)^2\bigg)
-\frac{\eta A_{\Sigma}z_t}{32ab^2G_N^{(4)}z_h^4}\int_{0}^{z_t}dz \theta(\upsilon) z^2\bigg(1-\big(\frac{z}{z_t}\big)^2\bigg),
\end{gather}
in which we use (3.28). By changing differential parameter to
$\upsilon$ and transformation of integral limit, there are two
different cases which could be happened by attention to the
saturation time. They are defined similar to the strip region as
$t_{sat}=z_t=a$: (a) for $t<t_{sat}$ the EE is a function of time
as
\begin{equation}
\Delta S(t<t_{sat})=\frac{\mu A_{\Sigma}}{16G_N^{(4)}z_t^2z_h^3}\big(2z_t^2t^2-t^4\big)
-\frac{\eta A_{\Sigma}}{480b^2G_N^{(4)}z_h^4z_t^2}\big(5z_t^2t^3-3t^5\big),
\end{equation}
and (b) for $t>t_{sat}$ the EE saturates to a fixed value as,
\begin{equation}
\Delta S(t>t_{sat})=\frac{ A_{\Sigma}}{16G_N^{(4)} }\big(\frac{z_t}{z_h}\big)^3\bigg(\frac{\mu}{z_t}-\frac{\eta}{15b^2z_h}\bigg).
\end{equation}
Similar to the strip case we first set $\Delta S_{sat}$ to be as
the saturated value of entanglement growth for $t>t_{sat}$ then we
can rewrite (3.31).
Diagrams of the evolution of EE in this case is plotted in figures (1.c), (2.c) and (1.d)
for different values of the parameters $b$ and $Q$. The result is very similar to the strip case and description is same as well.
 Figure (2.d) similar to the previous case indicates AdS-RN solution. \\
Also the instantaneous rate of entanglement growth defined in
(3.19) can be evaluated by noticing that in the ball region
considered here (2-dimension ball or a disk on the boundary)
$A_{\Sigma}=2\pi a$ and $V_A=\pi a^2$. The following relation
which is very similar to (3.20) would be obtained:
\begin{equation}
\Re(t)=\frac{a}{2\Delta S_{sat}}\frac{d(\Delta S(t))}{dt},
\end{equation}
where $\Delta S_{sat}$ which is defined in (3.32) includes the aspects of our gravity model, but doesn't play any role in this parameter.
Diagrams plotted in (3.c) and (3.d) represent the behavior of this function for ball region which is very similar to the strip case as qualitative but not quantitative.
For $Q=5$ as it plotted in (3.d), diagrams are very sensitive to the values of $b$, and like the strip region the value of $\Re_{max}$
 decreases by decreasing Born-Infeld parameter $b$, but when it goes to zero, $\Re_{max}$ changes its behavior and begins to increase until in $b=0.00155$
  crosses the speed of light (see dot-gray line). By these results we can see that the value of $b$ for which the rate of the evolution exceeds
  the speed of light happens just in smaller $b$ in the ball region relative to the strip case, and also as we observe in the figures (3.b) and (3.d),
  this exceeding the speed of light happens in the ball region earlier for the same $b$.
 \subsection{Thermalization after quench:} Similar to large
entangled regions given in refs. [14, 51], we can distinguish
three regimes in the evolution process of EE. At initial times
after turning on the quench we have a "\textit{pre-local
equilibration regime}". We can see for initial times $t\ll
t_{sat}=z_t$ both the strip and the ball regions have a same
behavior as follows.
\begin{equation}
\Delta S(t\ll t_{sat})=\frac{\mu
A_{\Sigma}}{8G_N^{(4)}z_h^3}~t^2+\cdots,
\end{equation}
in which dots denote to negligible higher order (small)
corrections. As we can see,
 the initial thermalization of EE has a universal behavior for global quantum quench.\\
In contrary with large subsystems we can't see a linear behavior
after the local equilibrium point, $z_h$, and so tsunami picture
breaks down here. But we can simulate this behavior for a specific
time $t_{max}$ at which the rate of entanglement growth is
maximum. By attention to (3.19) we can obtain the evolution close
to $t_{max}$ as a
 linear form as follows:\begin{equation}
\Delta S_A(t)=\Delta S_A(t_{max})+\Re_{max}s_{eq}A_{\Sigma}(t-t_{max})+\mathcal{O}(t-t_{max})^3,
\end{equation}
in which $s_{eq}$ is the EE density at $z_h$. Since there is not any quadratic term in the above equation,
 therefore the time behavior of the EE will be linear close to $t_{max}$.\\
As we can see from [14, 51], the behavior of the growth of EE just
before saturation for large subsystems depends on the shape of
entangled region. After saturation time the entanglement growth
ceases and the system reaches to an equilibrium state. This phase
transition depends on the shape of region, as for the strip case
happens suddenly (for dimensions larger than 3) which means a
first order transition, but not for ball regions. The latter case
is not occurred suddenly which means a second order phase
transition. The behavior of the system could be characterized by a
nontrivial scaling exponent $\gamma$ as follows:
\begin{equation}
\Delta S_A(t)\varpropto\Delta S_A(t_{sat})+(t_{sat}-t)^\gamma~~~~~\gamma=\frac{d+1}{2},
\end{equation}
in which $t$ is very close to saturation time. But in small
entangled regions phase transition is independent of the shape and
for all regions happens continuously (second order phase
transition). By expanding $\Delta S(t)$ around the saturation time
one can find critical exponent $\gamma=3/2$ in the leading order
for the strip, whereas it will be $\gamma=2$ for the ball region
similar to mean-field behavior.

\section{Conclusion}
 As a nonlocal observable we calculated HEE of a Born-Infeld AdS
 black hole for the entangled strip and the ball subregions on the holographic
 side. Our calculations restricted on small subregions and so we apply the perturbation method to obtain time dependent HEE.
 To do so we use step function time dependence for the black hole mass and charge. Applying any other profiles do not change physics of the problem.
 Our step function profile corresponds to a suddenly injecting
matter on the CFT side
 which reduces to a collapsing
 null shell on the bulk AdS spacetime. It makes an AdS Born Infeld black hole finally at
 center of the bulk
 AdS space time. We found out the saturation point will be depend on both electric charge $Q$ and Born-Infeld parameter $b$.
 The saturation
 entropy is increased by raising $b$ for a
  fixed charge, and when $b\rightarrow\infty$ it reduces to a simple AdS-RN black hole solution which
 is investigated in [47]. The behavior of system is independent of the shape of region
  and is same as for the strip and the ball regions.
 The instantaneous rate for both
 regions is studied and we see for very small $b$ it will be exceed the speed of light  just for large charges.
  Indeed when
 the electric charge has take some small values, then $\Re(t)$ dose not never exceed the speed of light. We also have a short
 investigation of various regimes which is plotted in diagrams and compare them with large subsystem case. Except the initial times, other
 regimes are different from large entangled region which is studied in [14, 51]. We can see phase transition to a saturated state always which
 continuous to small subregions with different critical exponent for the regions of the strip and the
 ball. Recently we calculated HEE of Gauss-Bonnet Black hole for which the results behave qualitatively similar to the results of the present work. In general Gauss-Bonnet black holes are obtained
 from higher order
 derivative gravity models [60] such as the 5D Lovelock gravity [52] where there is
 some topological invariants called as the
Gauss-Bonnet scalars
$R_{\alpha\beta\gamma\delta}R^{\alpha\beta\gamma\delta}-4R_{\alpha\beta}R^{\alpha\beta}+R^2$
[11, 53-59]. In fact the latter quantity is originated from the
renormalization of the expectation value of interacting quantum
fields stress tensor operator [4,5].
  \vskip .5cm
 \noindent
  {\bf References}
\begin{description}
\item[1.] J. M. Maldacena, "The Large N Limit of Superconformal Field Theories and Supergravity," Adv. Theor. Math. Phys. 2 (1998) 231;
 hep-th/9711200.
\item[2.] E. Witten, "Anti De Sitter Space And Holography," Adv. Theor. Math. Phys. 2 (1998) 253; hep-th/9802150.
\item[3.] G. 't Hooft, "Dimensional Reduction in Quantum Gravity," gr-qc/9310026.
\item[4.] N. D. Birrell and P. C. W. Davies, `Quantum fields in curved
space` (Cambridge, Cambridge University press, 1982)
\item[5.] L. Parker and B. Toms, "Quantum field theory in curved
spacetime" Cambridge, Cambridge University Press, (2009).
\item[6.] Y. Makeenko, "A Brief Introduction to Wilson Loops and Large
N", Phys. Atom. Nucl.73 ,878 (2010); hep-th/0906.4487.
\item[7.] V. Balasubramanian et al., "Thermalization of Strongly Coupled Field Theories," Phys.Rev. Lett. 106, 191601 (2011); hep-th/1012.4753.
\item[8.] V. Balasubramanian et al., "Holographic Thermalization," Phys. Rev. D 84, 026010(2011); hep-th/1103.2683.
\item[9.] D. Galante and M. Schvellinger, "Thermalization with a chemical potential from AdS
spaces" JHEP 1207, 096 (2012); hep-th/1205.1548.
\item[10.] E. Caceres and A. Kundu, "Holographic thermalization with chemical
potential" JHEP 1209, 055 (2012); hep-th/1205.2354.
\item[11.] X. X. Zeng and B. W. Liu, "Holographic thermalization in Gauss-Bonnet
gravity", Phys. Lett. B729, 481 (2013); hep-th/1305.4841.
\item[12.] X. X. Zeng, X. M. Liu and B. W. Liu, "Holographic thermalization with a chemical potential in Gauss-Bonnet
gravity" JEHP 03, 031 (2014); hep-th/1311.0718.
\item[13.] X. X. Zeng, D.Y.Chen and L. F. Li, "Holographic thermalization and gravitational collapse in the spacetime dominated by quintessence dark
energy" Phys. Rev. D91, 046005 (2015); hep-th/1408.6632.
\item[14.] H. Liu and S. J. Suh, "Entanglement Tsunami: Universal Scaling in holographic
thermalization" Phys. Rev. Lett.112, 011601 (2014);
hep-th/1305.7244.
\item[15.] S. J. Zhang and E. Abdalla, "Holographic thermalization in charged Dilaton Anti-de Sitter space
time" Nucl. Phys. B896, 569 (2015); hep-th/1503.07700.
\item[16.] A. Buchel, R. C. Myers and A. V. Niekerk, "Nonlocal probe of thermalization in holographic quenches with spectral
methods" JHEP 02, 017 (2015); hep-th/1410.6201.
\item[17.] B. Craps,  E. Kiritsis, C. Rosen, A. Taliotis, J. Vanhoof and H. Zhang, "Gravitational collapse and thermalization in the hard wall
model", JHEP 02, 120 (2014); hep-th/1311.7560.
\item[18.] T. Albash and  C. V. Johnson, "Holographic Studies of Entanglement Entropy in
Superconductors",JHEP 1205, 079 (2012); hep-th/1202.2605.
\item[19.] R. G. Cai, S. He, L. Li and Y. L. Zhang, "Holographic Entanglement Entropy in Insulator/Superconductor
Transition", JHEP 1207, 088 (2012), hep-th/1203.6620
\item[20.] R. G. Cai, L. Li, L. F. Li and R. K. Su, "Entanglement Entropy in Holographic P-Wave Superconductor/Insulator
Model", JHEP 1306,  063, (2013); hep-th/ 1303.4828
\item[21.] L. F. Li, R. G. Cai, L. Li and C. Shen, "Entanglement entropy in a holographic p-wave superconductor
model" Nucl. Phys. B894, 15 (2015); hep-th/1310.6239
\item[22.] R. G. Cai, S. He, L. Li, Y. L. Zhang, "Holographic Entanglement Entropy in Insulator/Superconductor
Transition",JHEP 1207,  088 (2012); hep-th/1203.6620
\item[23.] X. Bai, B. H. Lee, L. Li, J. R. Sun and H. Q. Zhang, "Time Evolution of Entanglement Entropy in Quenched Holographic
Superconductors", JHEP 04,066 (2015); hep-th/1412.5500
\item[24.] R. G. Cai, L. Li, L. F. Li and R. Q. Yang, "Introduction to Holographic Superconductor
Models", Sci. China-Phys. Mech. Astron, 58,(6),060401 (2015);
hep-th/1502.00437
\item[25.] Y. Ling, P. Liu, C. Niu, J. P. Wu, Z. Y. Xian, "Holographic Entanglement Entropy Close to Quantum Phase
Transitions", JHEP, 04, 114 (2016); hep-th/1502.03661
\item[26.] S. A. Hartnoll, ""Lectures on holographic methods for condensed matter physics,â€" Class. Quant.
Grav. 26, 224002 (2009); hep-th/0903.3246.
\item[27.] N. Engelhardt, T. Hertog and  G. T. Horowitz, "Holographic Signatures of Cosmological
Singularities", Phys. Rev. Lett. 113, 121602 (2014);
hep-th/1404.2309
\item[28.] N. Engelhardt, T. Hertog and G. T. Horowitz, "Further Holographic Investigations of Big Bang
Singularities",     JHEP 1507,  044 (2015); hep-th/1503.08838
\item[29.] S. Ryu and T. Takayanagi, "Holographic derivation of entanglement entropy from AdS/CFT,"
Phys. Rev. Lett. 96, 181602 (2006); hep-th/0603001.
\item[30.] R. Emparan, `Black hole entropy as entanglement entropy: a
holographic derivation` JHEP 0606, 012 (2006) ; hep-th/0603081
\item[31.] V. E. Hubeny, M. Rangamani, and T. Takayanagi, "
A Covariant holographic entanglement entropy proposal", JHEP 0707,
 062 (2007), hep-th/0705.0016.
\item[32.] U. H. Danielsson, E. Keski-Vakkuri and M. Kruczenski, "Spherically collapsing matter
in AdS, holography and shellons," Nucl. Phys. B 563, 279 (1999);
hep-th/9905227.
\item[33.] U. H. Danielsson, E. Keski-Vakkuri and M. Kruczenski, "Black hole formation in AdS and thermalization on the boundary," JHEP 0002,
 039 (2000); hep-th/9912209.
\item[34.] S. B. Giddings and A. Nudelman, "Gravitational collapse and its boundary description in AdS," JHEP 0202, 003 (2002); hep-th/0112099.
\item[35.] P. Calabrese and J. L. Cardy, "Evolution of entanglement entropy in one-dimensional systems," J. Stat. Mech. 0504, 04010 (2005);
cond-mat/0503393.
\item[36.] H. Casini, H. Liu and M. Mezei, "Spread of entanglement and causality," JHEP 1607,077 (2016); hep-th/1509.05044.
\item[37.] T. Hartman and N. Afkhami-Jeddi, "Speed Limits for Entanglement", hep-th/1512.02695.
\item[38.] G. Camilo, B. Cuadros-Melgar, E. Abdalla, "
Holographic thermalization with a chemical potential from
Born–Infeld electrodynamics", JHEP. 02, 103 (2015)'
hep-th/1412.3878.
\item [39.] X. X. Zeng, X. M. Liu and L. F. Li, "
Phase structure of the Born-Infeld-anti-de Sitter black holes
probed by non-local observable" Eur. Phys. J. C  76, 616
  (2016); hep-th/1601.01160
\item[40.] R. G. Cai, D. W. Pang, A. Wang, "Born-Infeld Black Holes in (A)dS Spaces", Phys. Rev. D 70, 124034 (2004);
hep-th/0410158.
\item[41.] T. K. Dey, "Born-Infeld black holes in the presence of a cosmological constant", Phys. Lett. B 595, 484 (2004); hep-th/0406169.
\item[42.] T.K. Dey, Born-Infeld
black holes in the presence of a cosmological constant, Phys.
Lett. B595, 484 (2004) hep-th//0406169
\item[43.] E. Witten, "Anti-de Sitter Space, Thermal Phase Transition, And Confinement In Gauge Theories", Adv. Theor. Math.
Phys. 2,  505 (1998); hep-th/9803131.
\item[44.] A. Chamblin, R. Emparan, C. V. Johnson and R. C. Myers, ""Charged AdS black holes and catastrophic holography,â€"
 Phys. Rev. D60, 064018,(1999); hep-th/9902170.
\item[45.] H. Ghaffarnejad, "Quantum field
backreaction corrections and remnant stable evaporating
Schwarzschild-de Sitter dynamical black hole", Phys. Rev. D75,
084009 (2007)
\item[46.] H. Ghaffarnejad, H. Neyad and M. A. Mojahedi, "
Evaporating quantum Lukewarm black holes final state from
backreaction corrections of quantum scalar fields", Astrophys.
Space Sci, 346, 497 (2013);  physics.gen-ph/1305.6914.
\item[47.] S. Kundu and J. F. Pedraza, "Spread of entanglement for small subsystems in holographic CFTs,"
Phys. Rev. D 95, 086008 (2017); hep-th/1602.05934
\item[48.] F. Moura, "Absorption of scalars by extremal black holes in string theory",Gen. Relativ. Gravit. 49,  117 (2017); hep-th/1406.3555.
\item[49.]  E. Caceres, M. Sanchez and J. Virrueta, "
"Holographic Entanglement Entropy in Time Dependent Gauss-Bonnet
Gravity,â€" hep-th/1512.05666.
 \item[50.] V. E. Hubeny, "Extremal surfaces as bulk probes in AdS/CFT", JHEP 07, 093 (2012); hep-th/1203.1044
\item[51.] H. Liu and S. J. Suh, "Entanglement growth during thermalization in holographic systems," hep-th/1311.1200.
\item[52.] J. de Boer, M. Kulaxizi, and A. Parnachev, Holographic Entanglement Entropy in Lovelock Gravities, JHEP 1107, 109, (2011); hep-th/1101.5781.
\item[53.] L. Y. Hung, R. C. Myers, and M. Smolkin, "On Holographic Entanglement Entropy and Higher Curvature Gravity"
, JHEP 1104, 025 (2011), hep-th/1101.5813.
\item[54.] X. Dong, "Holographic Entanglement Entropy for General Higher Derivative Gravity", JHEP 1401, 044, (2014); hep-th/1310.5713.
\item[55.] J. Camps, "Generalized entropy and higher derivative Gravity", JHEP 1403, 070, (2014); hep-th/1310.6659.
\item[56.] S. Hansraj, "Generalized spheroidal spacetimes in 5D Einstein-Maxwell-Gauss-Bonnet gravity" Eur. Phys. J. C, 77, 557 (2017).
\item[57.] R. G. Cai, "Gauss-Bonnet Black Holes in AdS Spaces", Phys. Rev. D 65, 084014 (2002); hep-th/0109133
\item[58.]  E. Caceres, M. Sanchez and J. Virrueta, ""Holographic Entanglement Entropy in Time Dependent Gauss-Bonnet Gravity,â€"
hep-th/1512.05666.
\item[59.] S. He, L. F. Li and X.X. "Zeng, Holographic Van der Waals-like phase transition in the Gauss-Bonnet gravity," Nucl. Phys. B915, 243, 261 (2017);
hep-th/1608.04208.
\item[60.] H. Ghaffarnejad, E. Yaraie, and M. Farsam. "Holographic Thermalization in AdS-Gauss-Bonnet gravity for Small entangled Regions." arXiv preprint arXiv:1806.05976 (2018).
\item[61.] E. Caceres, A. Kundu, J. F. Pedraza and D. L. Yang. "Weak field collapse in AdS: introducing a charge density." Journal of High Energy Physics 2015, no. 6 (2015): 111.
\end{description}
\begin{figure}[ht]
\centering
\subfigure[{}]{\label{1}
\includegraphics[width=.45\textwidth]{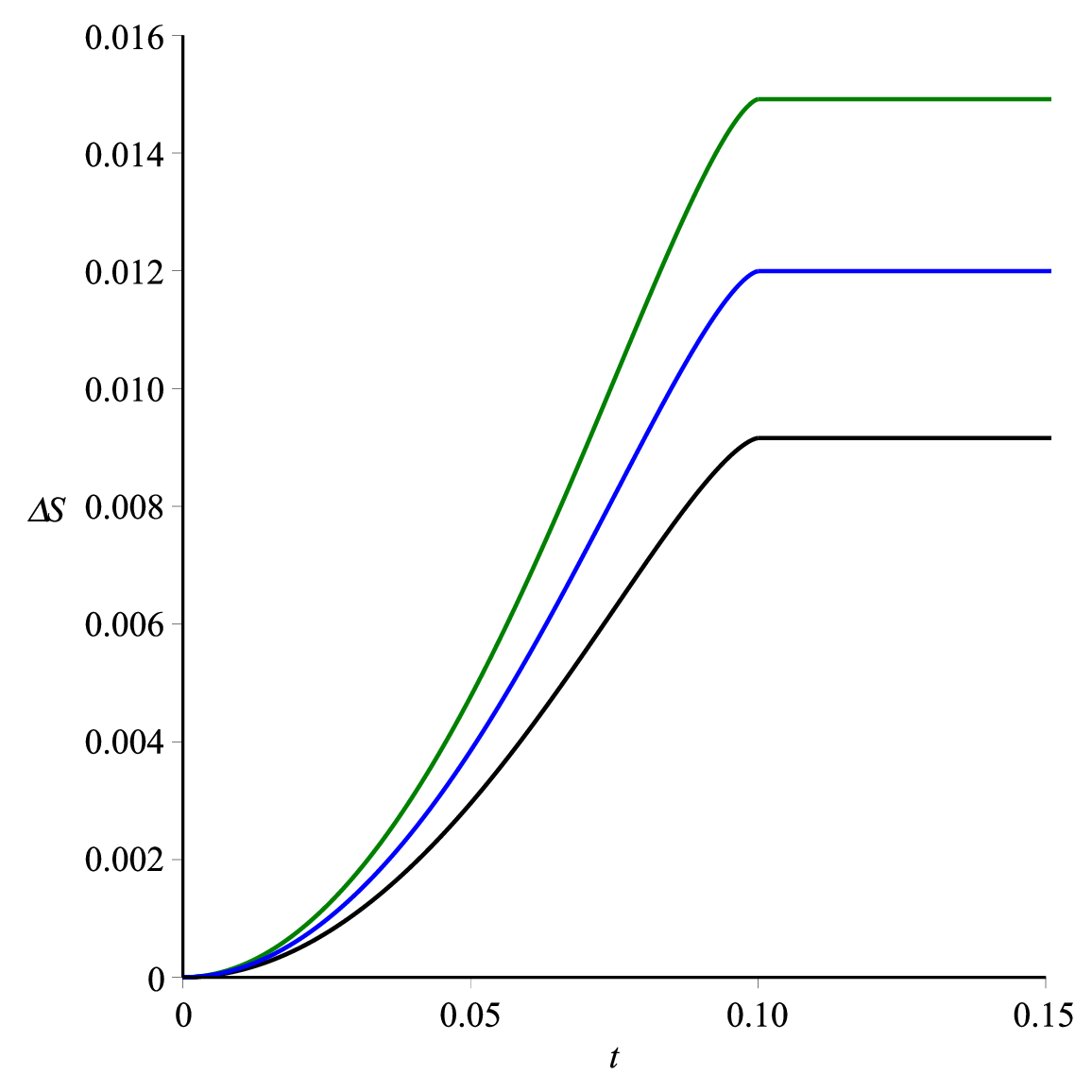}}
\hspace{3mm}
\subfigure[{}]{\label{1}
\includegraphics[width=.45\textwidth]{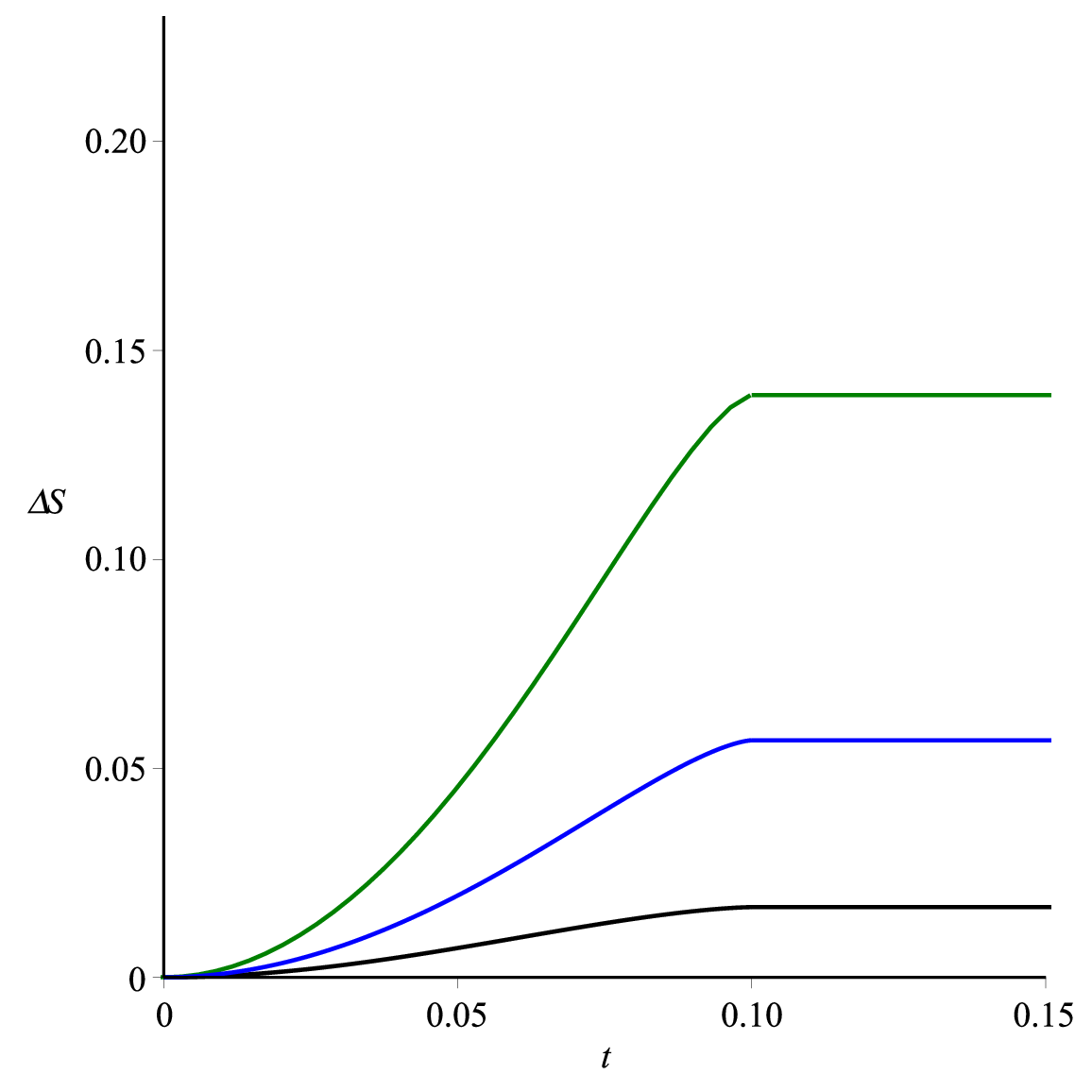}}
\hspace{3mm}
\subfigure[{}]{\label{1}
\includegraphics[width=.45\textwidth]{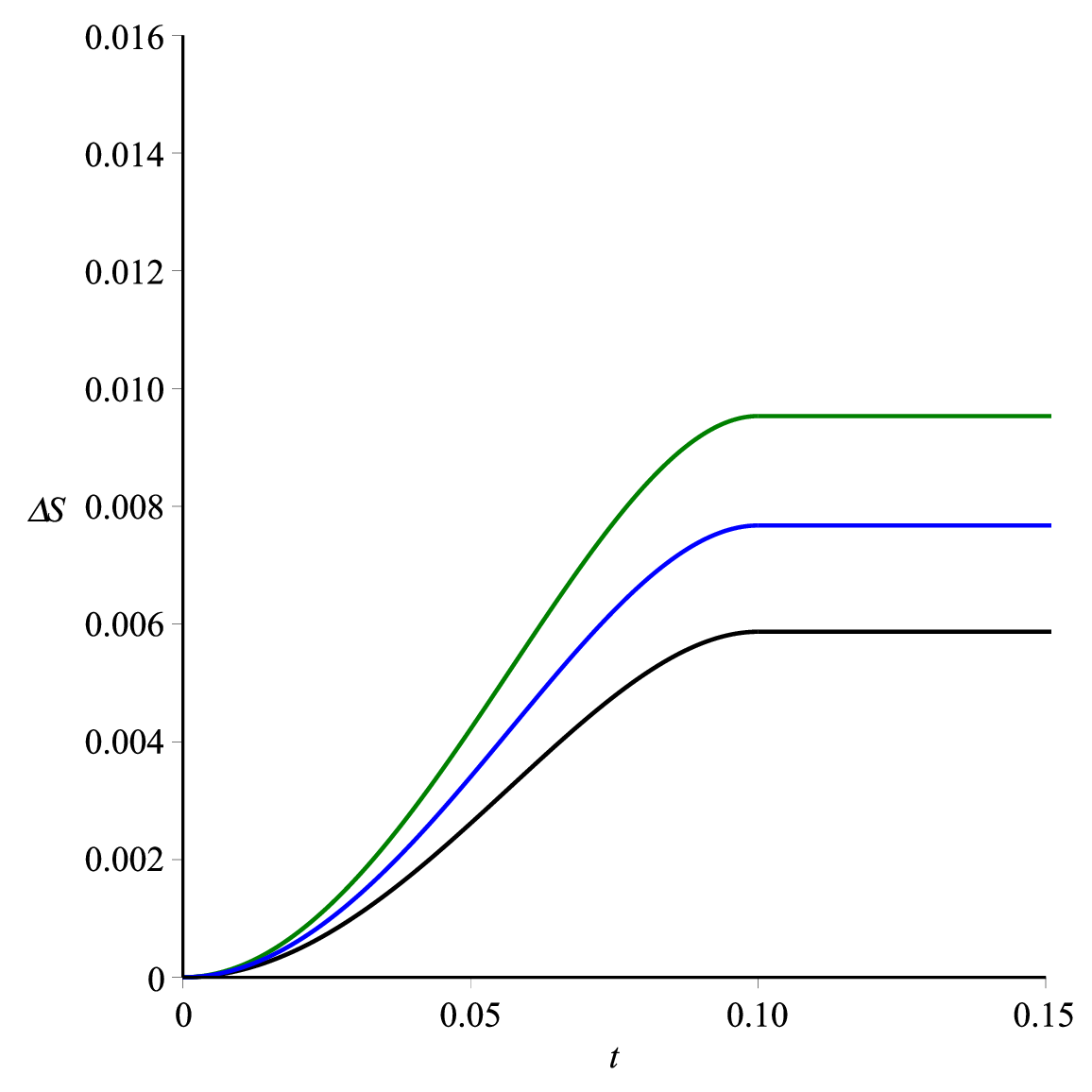}}
\hspace{3mm}
\subfigure[{}]{\label{1}
\includegraphics[width=.45\textwidth]{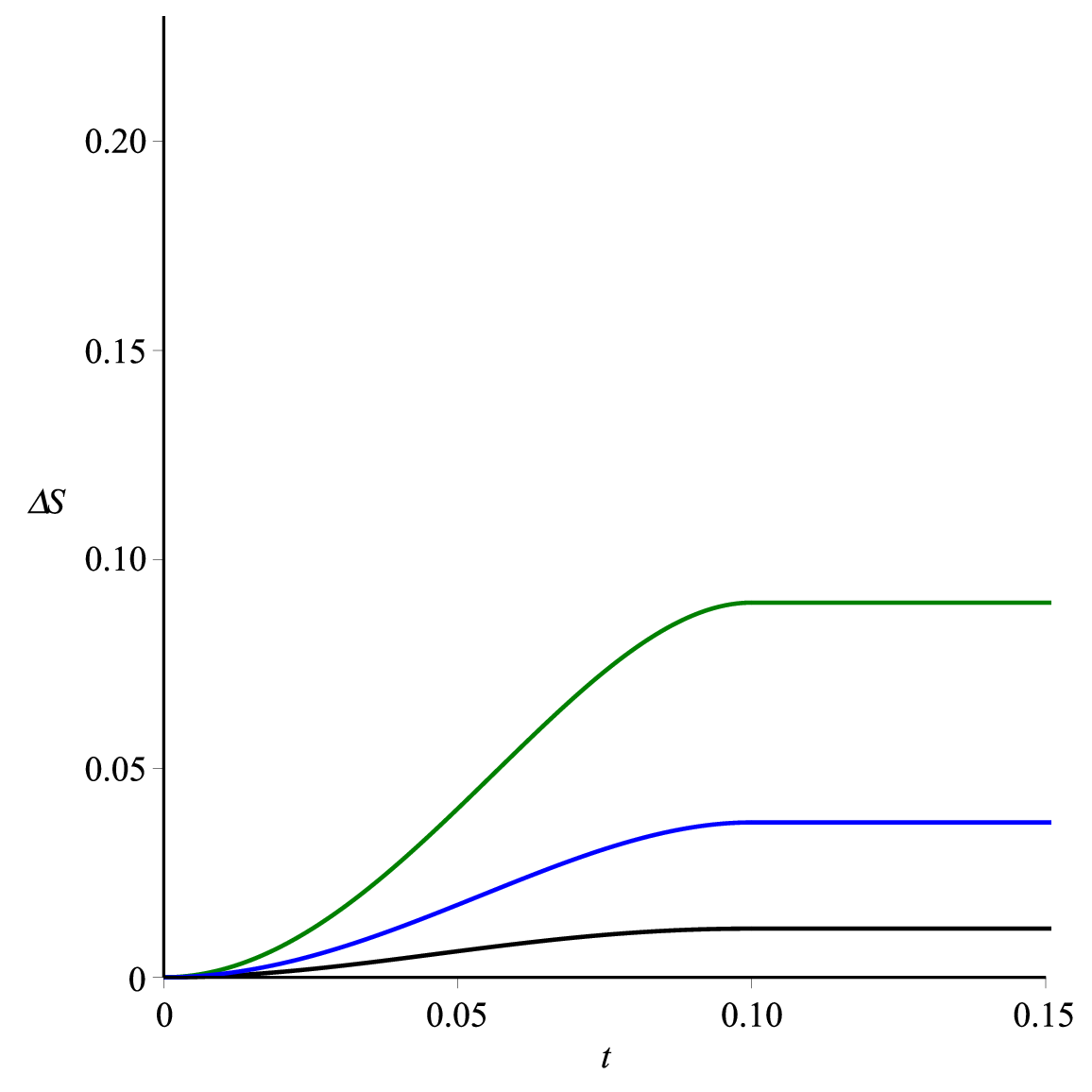}}
\caption{ Growth of EE is plotted vs the time for the strip and
the ball regions. In these diagrams we put $z_t/z_h=0.1$ due to
small subregions and set $A_{\Sigma}/16G_N^{(4)}=1$. In $(a)$ and
$(b)$ diagrams are plotted for the strip region and in $(c)$ and
$(d)$ are plotted for disk.  In $(a)$ and $(c)$ we set $Q=1$ but
in $(b)$ and $(d)$ we set $Q=5$. Black, blue and green colored
lines indicate $b=0.01, 0.1$ and $1$ respectively.} \label{l}
\end{figure}

\begin{figure}[ht]
\centering
\subfigure[{}]{\label{1}
\includegraphics[width=.45\textwidth]{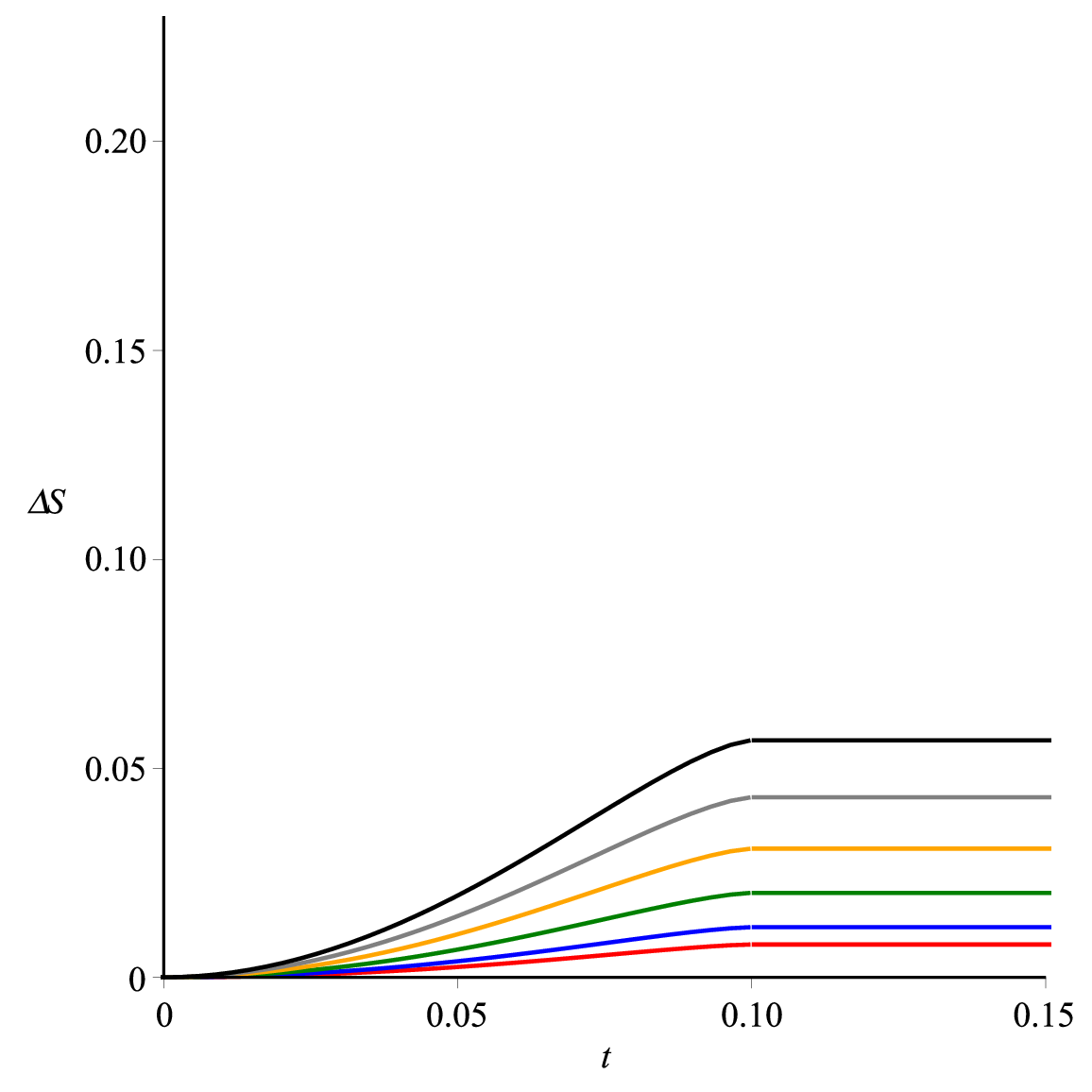}}
\hspace{3mm}
\subfigure[{}]{\label{1}
\includegraphics[width=.45\textwidth]{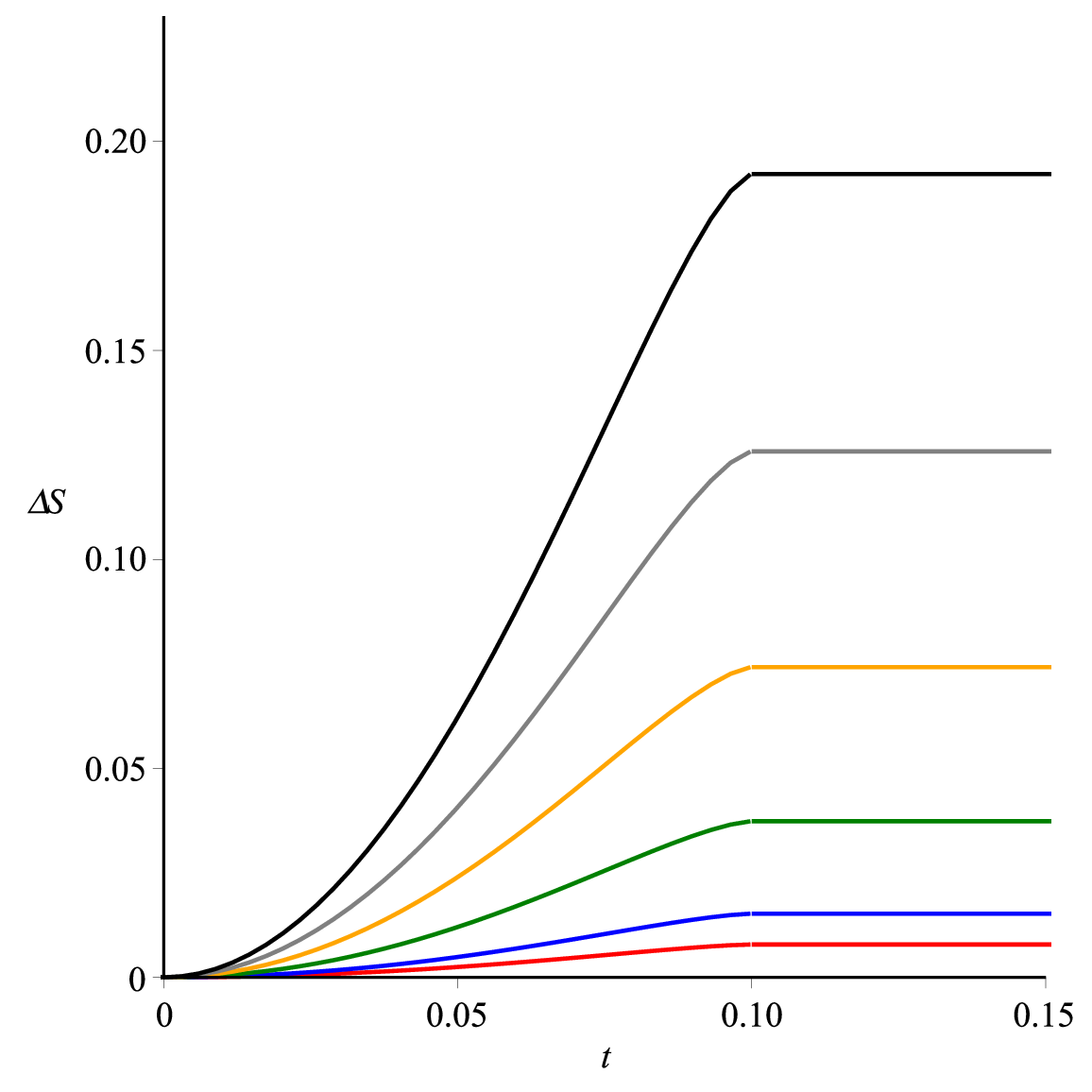}}
\hspace{3mm}
\subfigure[{}]{\label{1}
\includegraphics[width=.45\textwidth]{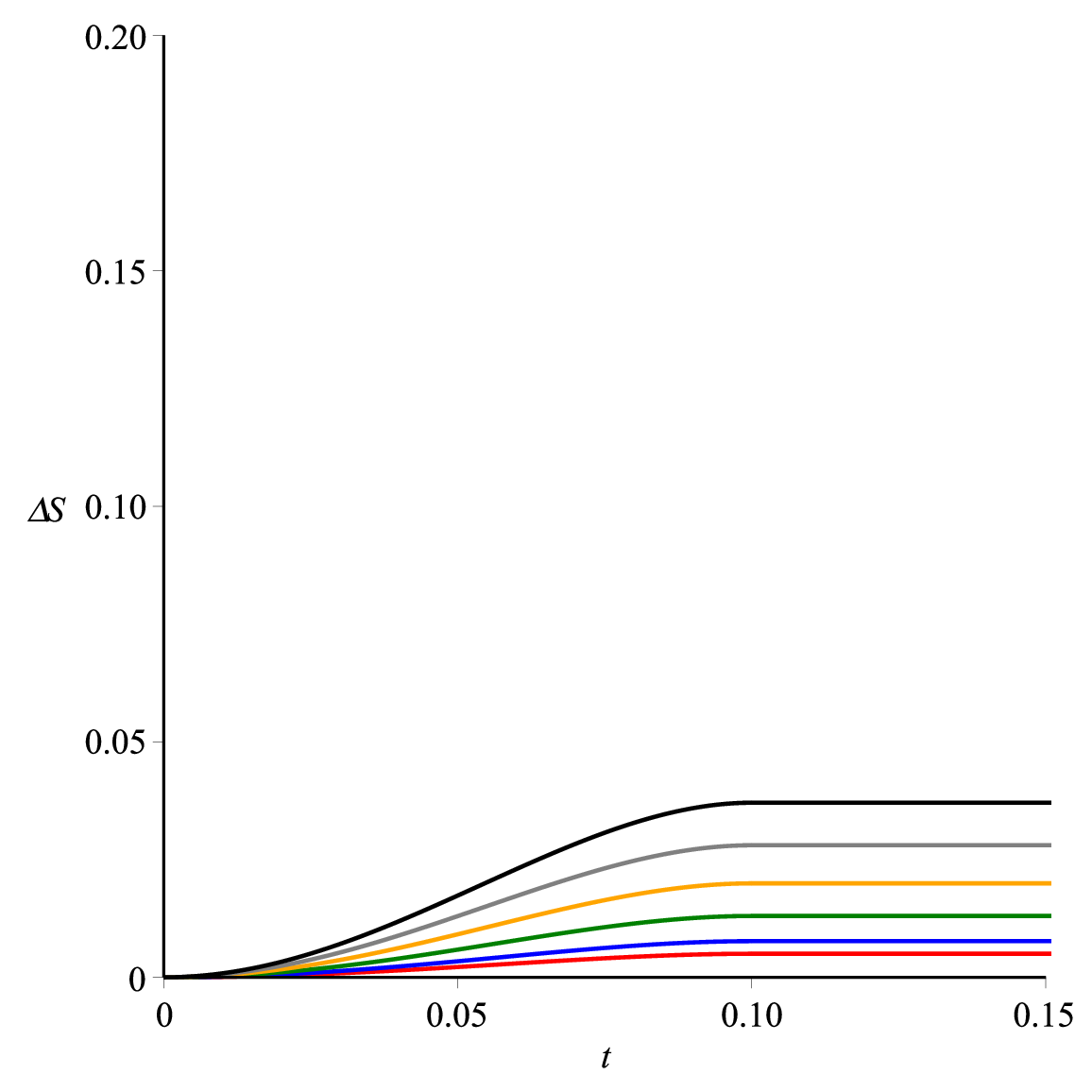}}
\hspace{3mm}
\subfigure[{}]{\label{1}
\includegraphics[width=.45\textwidth]{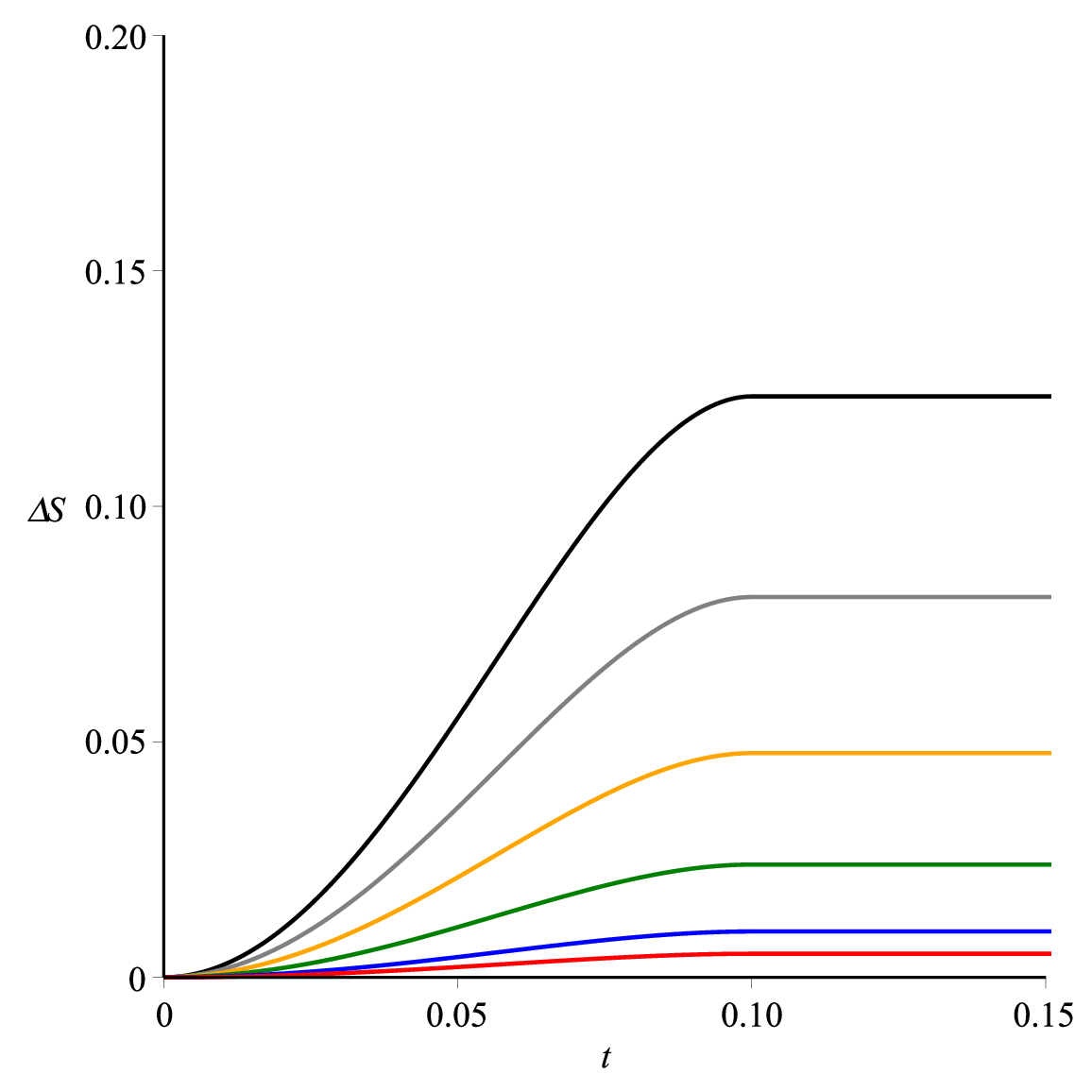}}
\caption{The evolution of EE for fixed $b=0.1$ and in the limit of
$b\rightarrow\infty$ for the strip region indicated in $(a)$ and
$(b)$, and for the ball region indicated in $(c)$ and $(d)$,
respectively. Colored lines represent different charges as
$Q=0,1,2,3,4,5$ which are indicated by red, blue, green, orange,
gray and black colored lines respectively. } \label{l}
\end{figure}

\begin{figure}[t]
\centering
\subfigure[{}]{\label{1}
\includegraphics[width=.45\textwidth]{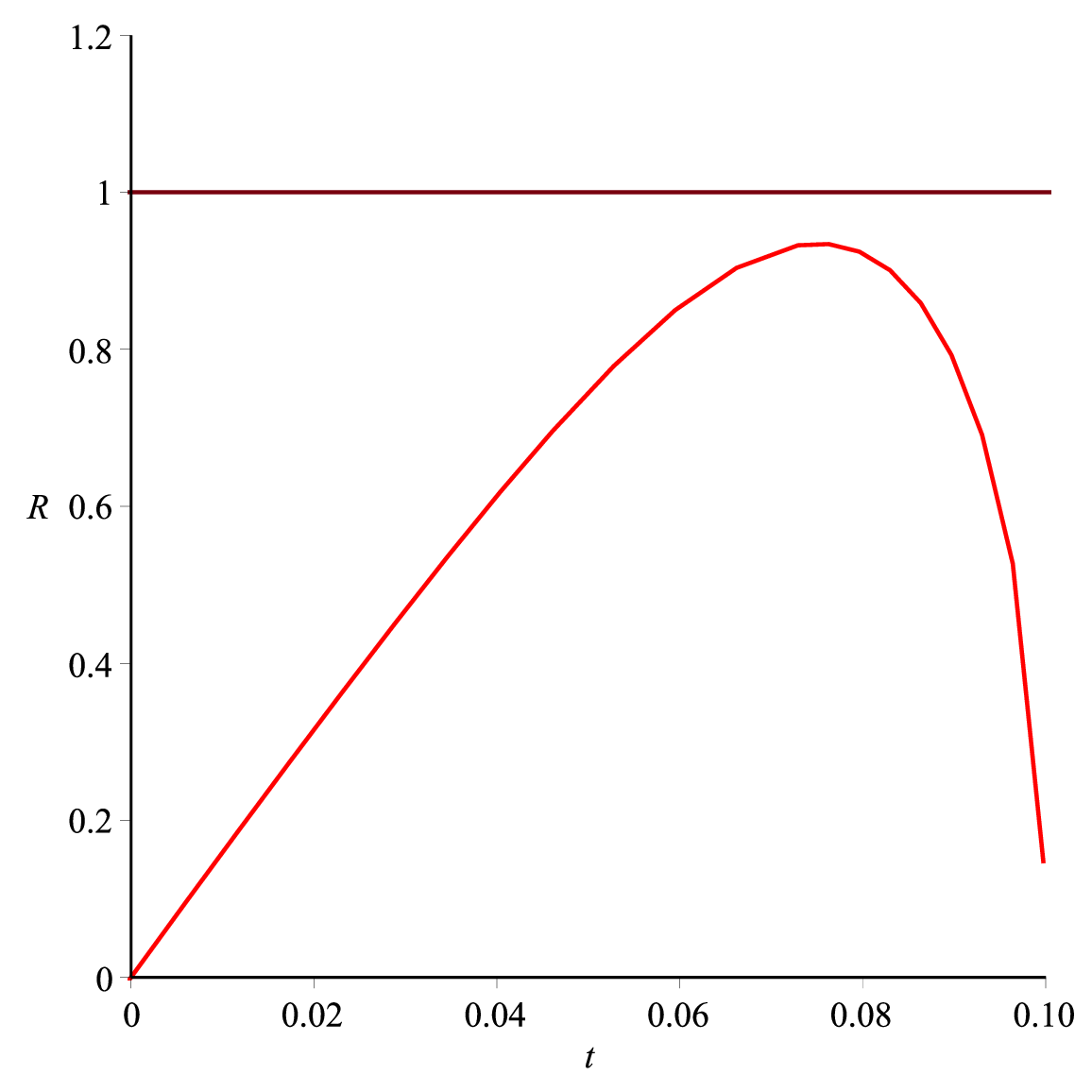}}
\hspace{3mm}
\subfigure[{}]{\label{1}
\includegraphics[width=.45\textwidth]{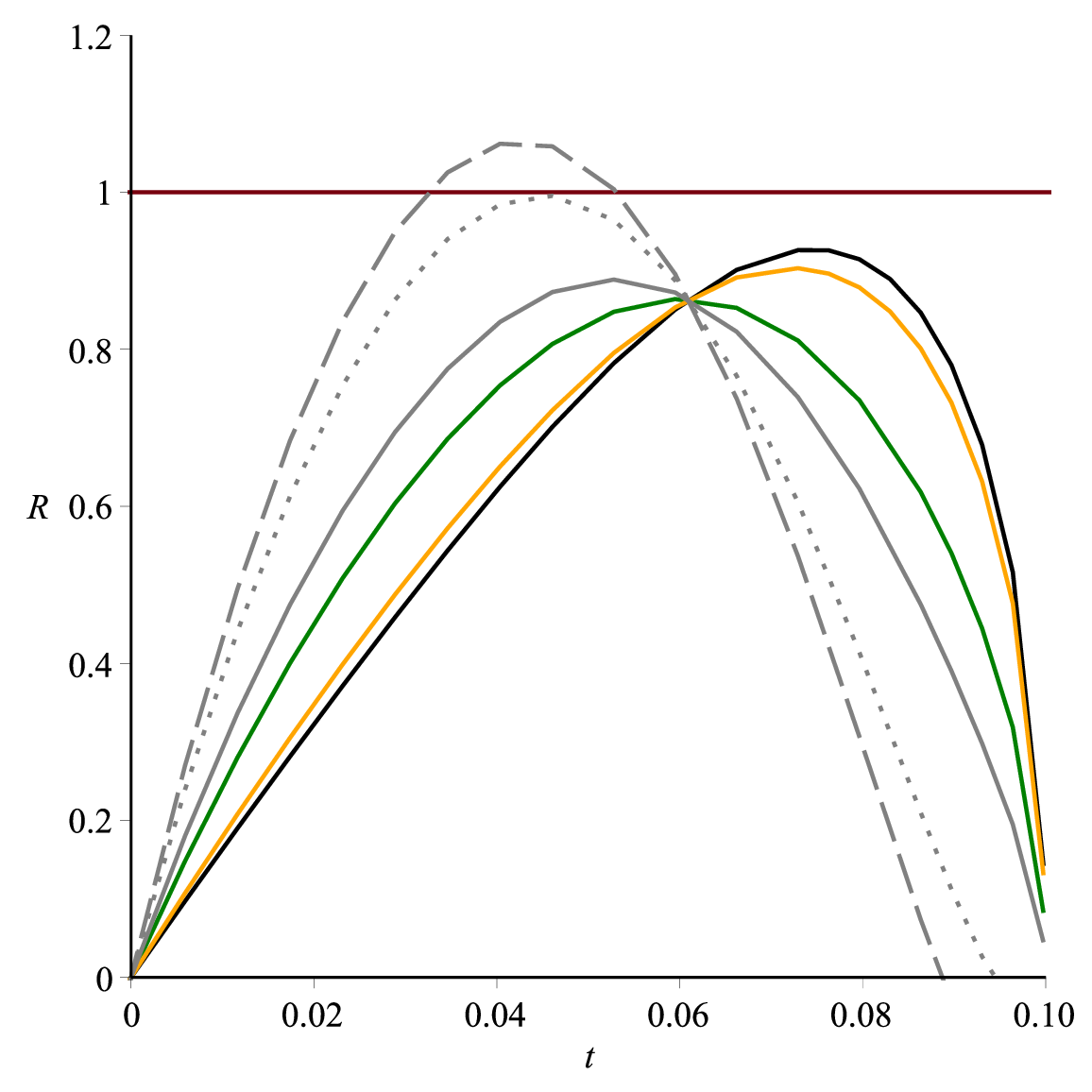}}
\hspace{3mm}
\subfigure[{}]{\label{1}
\includegraphics[width=.45\textwidth]{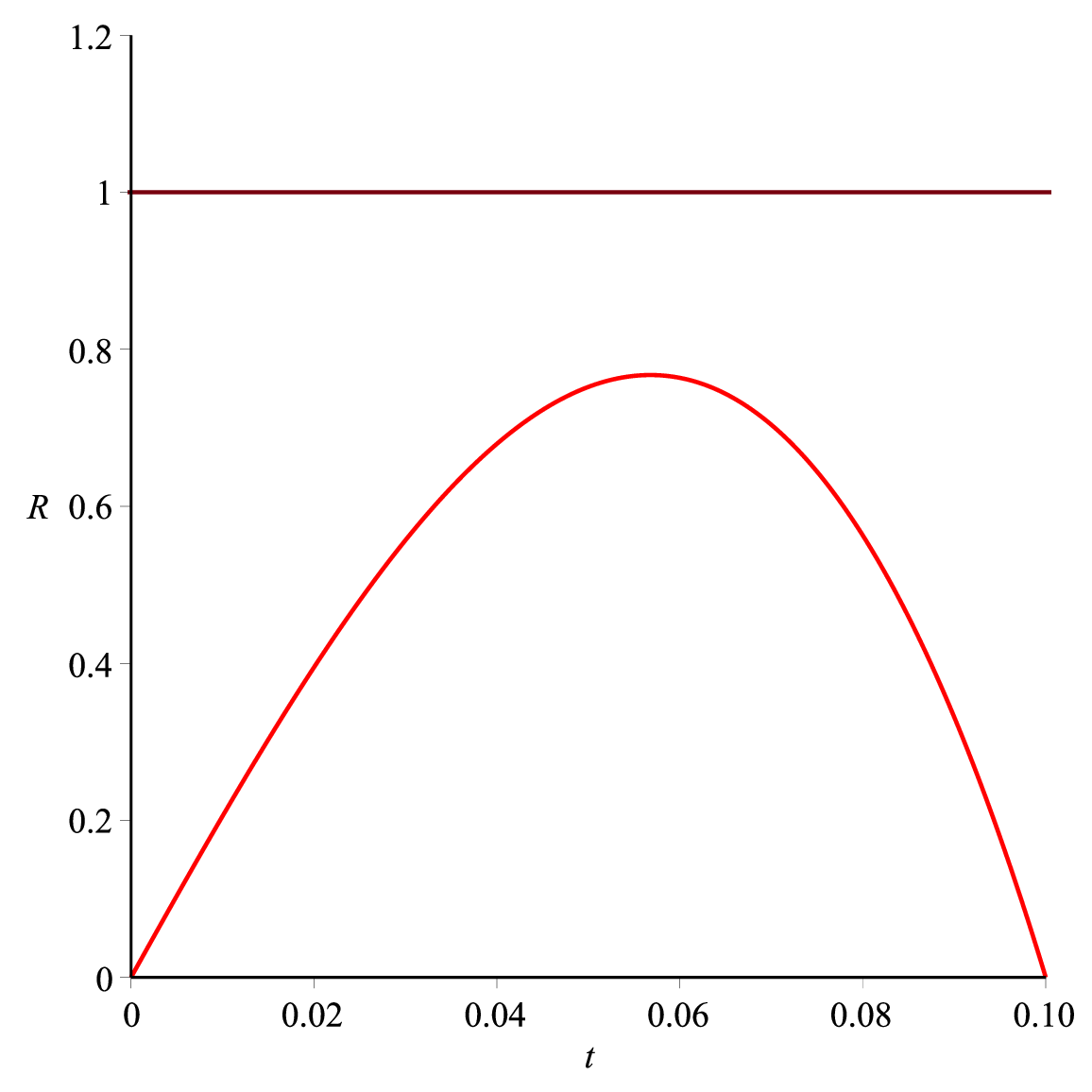}}
\hspace{3mm}
\subfigure[{}]{\label{1}
\includegraphics[width=.45\textwidth]{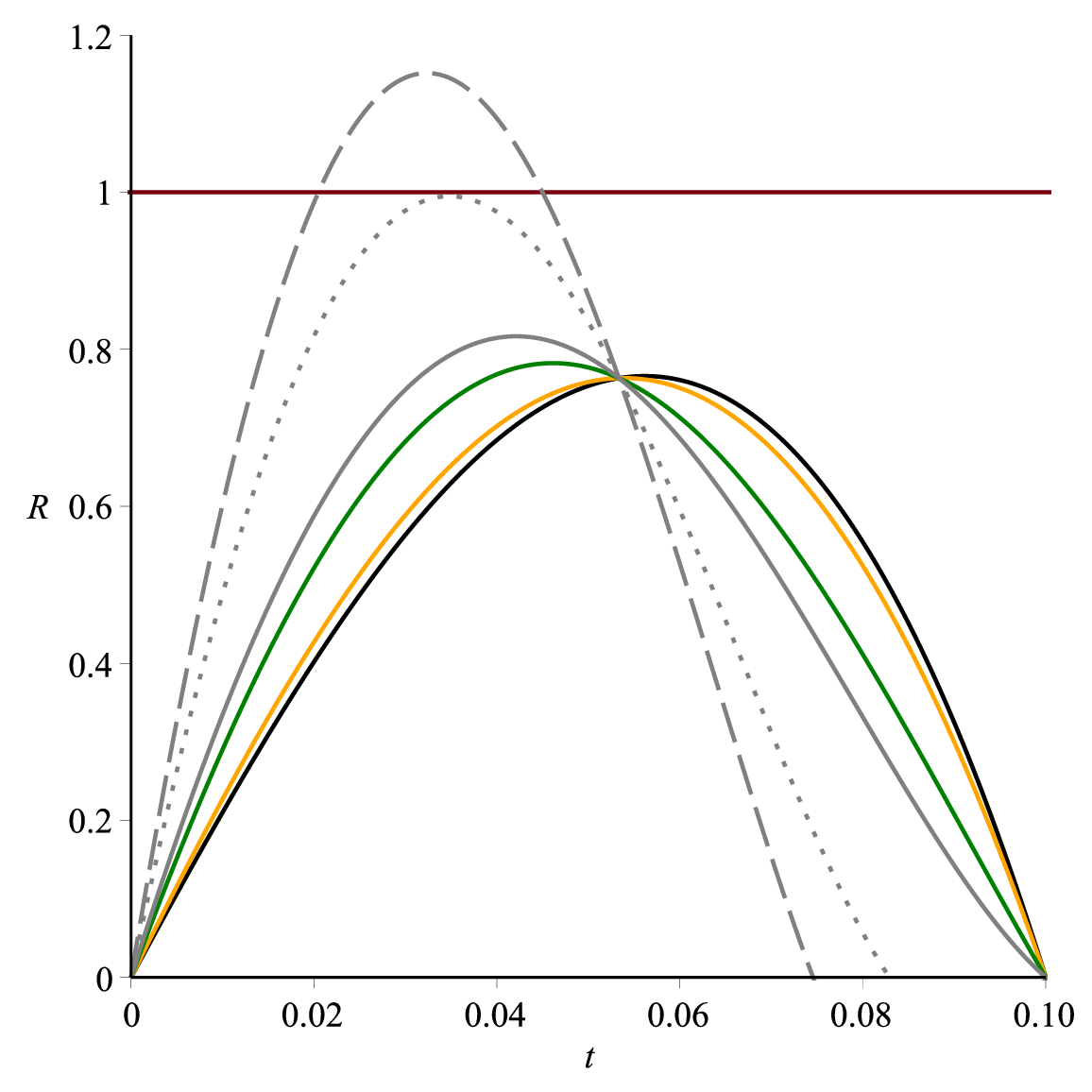}}
\caption{  The rate of EE growth for $z_t=z_h=1$, horizontal lines
represent the speed of light.(a) and (b) correspond to the strip
region while (c) and (d) correspond to the disk region. In figures
$(a)$ and $(c)$ we put $Q=1$ and plot  $\Re(t)$ vs the time for
all values of $b$ with red colored line. But in figures $(b)$ and
$(d)$ we put $Q=5$ and $b=1, 0.1, 0.01$ and $0.005$ and plot
$\Re(t)$ vs the time with black, orange, green and gray colored
lines respectively. In $(b)$ gray-dot line corresponds to
$b=0.0025$ and gray-dash line is obtained with $b=0.002$, whereas
in $(d)$ $b=0.00155$ corresponds to the dot-gray and $b=0.001$ is
used to plot dash-gray line. } \label{l}
\end{figure}

\end{document}